\documentclass[
twocolumn,
amsmath,amssymb,
aps,
physrev,
]{revtex4-2}

\usepackage{graphicx}
\usepackage{dcolumn}
\usepackage{bm}
\begin{document}

\title{\textbf{Universal Linear Response of First-Passage Kinetics: \\A Framework for Prediction and Inference} 
}% 

\author{Tommer D. Keidar}
\affiliation{%
School of Chemistry, The Center for Computational Molecular and Materials Science, The Center for Physics and Chemistry of Living Systems, Tel Aviv University, Tel Aviv, Israel
}%
 
\author{Shlomi Reuveni}%
 \email{Contact author: shlomire@tauex.tau.ac.il}
\affiliation{%
School of Chemistry, The Center for Computational Molecular and Materials Science, The Center for Physics and Chemistry of Living Systems, Tel Aviv University, Tel Aviv, Israel
}%

\date{\today}

\begin{abstract}
First-passage processes are pervasive across numerous scientific fields, yet a general framework for understanding their response to external perturbations remains elusive. While the fluctuation-dissipation theorem offers a complete linear response theory for systems in steady-state, it fails to apply to transient first-passage processes. We address this challenge by focusing on \textit{rare}---rather than weak---perturbations. Surprisingly, we discover that the linear response of the mean first-passage time (MFPT) to such perturbations is universal. It depends solely on the first two moments of the unperturbed first-passage time and the mean completion time following perturbation activation, without any assumptions about the underlying system's dynamics. To demonstrate the utility of our findings, we analyze the MFPT response of drift-diffusion processes in two scenarios: (i) stochastic resetting with information feedback, and (ii) an abrupt transition from a linear to a logarithmic potential. In both cases, our approach bypasses the need for explicit determination of the perturbed dynamics, unraveling a highly non-trivial response landscape with minimal effort.  Finally, we show how our framework enables a new type of experiment—inferring molecular-level fluctuations from bulk measurements, a feat previously believed to be impossible. Overall, the newly discovered universality reported herein offers a powerful tool for predicting the impact of perturbations on kinetic processes — and, remarkably, for extracting hidden single-molecule fluctuations from accessible bulk measurements.
\end{abstract}

\maketitle

\section{Introduction}
Linear response theory and fluctuation-dissipation relations stand as fundamental pillars in modern statistical mechanics \cite{FDR_review}. They provide a critical link between the inherent fluctuations within a system and its reaction to a subtle perturbation \cite{FDR_review}. One noteworthy illustration of this concept is the Einstein-Smoluchowski relation, which establishes a connection between the position fluctuations of an overdamped particle (reflected in the diffusion coefficient) and the response of its velocity to a weak force (represented by the mobility) \cite{Einstein_BM}. Another well-established example is the Johnson–Nyquist noise, elucidating the relationship between current fluctuations in an electrical circuit and the current's response to a weak voltage (captured by the conductivity) \cite{Johnson, Nyquist}. 

Classical linear response theory is limited to systems that are perturbed from their steady state. Consequently, it does not apply to transient processes, with first-passage processes serving as a notable example.
First-passage processes are a class of stochastic processes in which the question of interest is when a system will reach a specific state for the first time, i.e., the first-passage time (FPT) \cite{redner_book, FP_review_Bray_Majumdar_Scher, FP_and_applications, Target_Search_Problems}. First-passage processes play a central role in physics \cite{LRW_in_arb_dimensions, Stochastic_BD, Non_Markov_MFPT, Run_with_brown, condamin2007,Energy_landscape_FPT}, chemistry \cite{Diffusing_Diffusivities_limted_reactions, Energy_landscape_FPT, Sokolov2003Cyclization, Collect_realse, Sticky_particle}, biology \cite{Search_on_DNA, Transport_on_fractal_networks, Circular_DNA_Search, n_steps_memory, Spat_inhom_search, FP_in_cell_biology, Chemical_kinetics_beyond}, and finance \cite{JPBouchaud_book_chapter}. Some examples are the time taken to: cross a potential barrier \cite{Chupeau_potential_shape, Hanggi_Kramers}, for two molecules to meet and react \cite{Diffusion_and_surface_phenomena, Molecular_Kinetics_cond_phase, Nature_chem_benichou}, for a stock to reach a certain target price, for a stochastic optimization algorithm to converge, for specie to extinct \cite{Asaf2010, Ovaskainen2010}, and for an avalanche to start or end \cite{Colaiori2008}. The broad spectrum of applications underlines the importance of general results and unifying statements that apply to all first-passage processes.

\begin{figure}[t]
    \centering
    \includegraphics[width=0.4\textwidth]{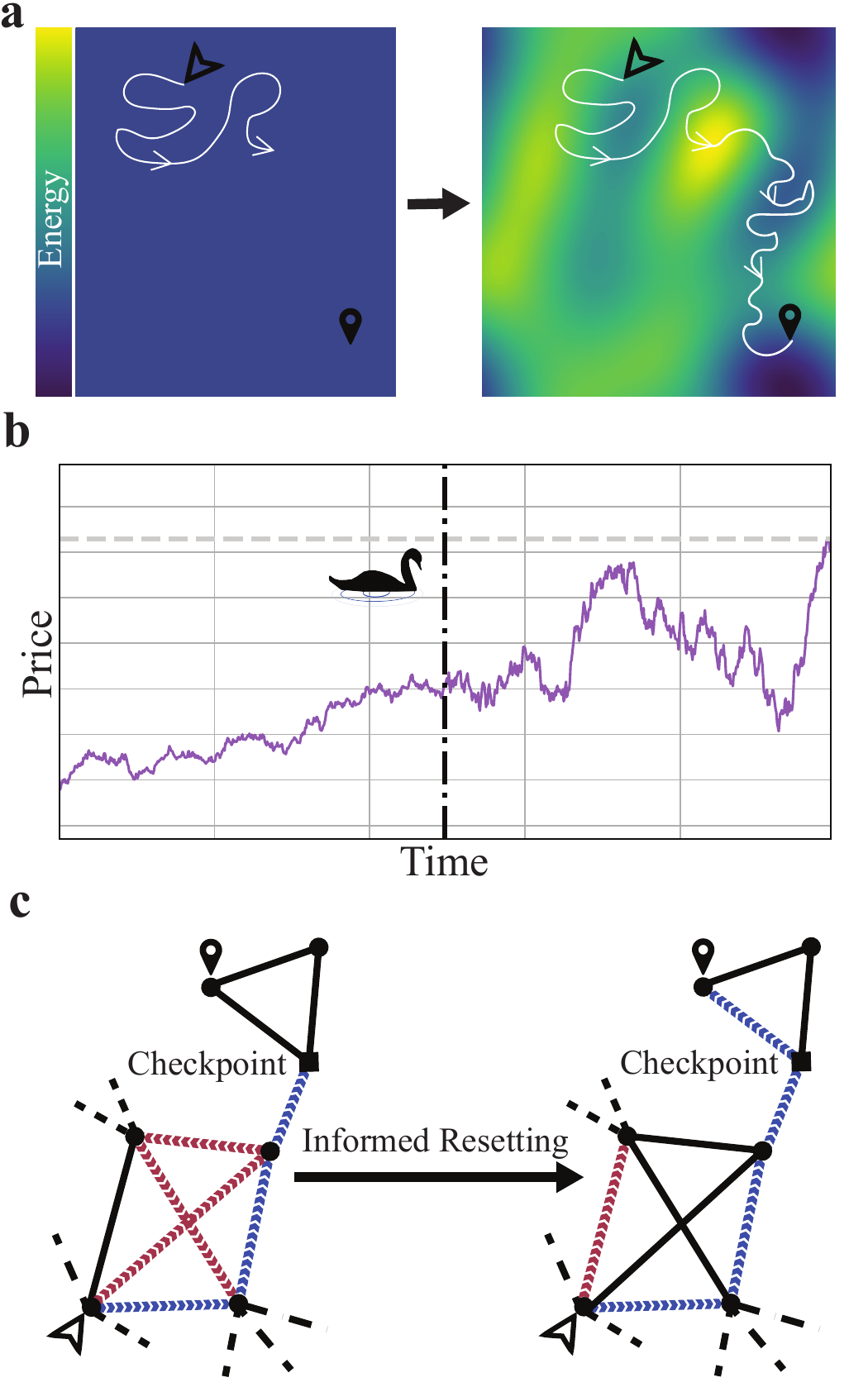}
    \caption{Examples of perturbed first-passage processes. \textbf{a}, A particle freely diffuses in search of a target. At some point in time, a field is activated and the particle continues to diffuse on the new potential. \textbf{b}, A stock is designated to be sold at a certain target price. A black swan changes market conditions, affecting the time at which the sell price is reached. \textbf{c}, An algorithm conducts a random search on a network. After passing a specific checkpoint (blue trajectory), the search is bound to nodes in the neighborhood of the solution. Trajectories that did not pass the checkpoint are stochastically restarted (red trajectory).}
    \label{fig:pertuebations}
\end{figure}

First-passage processes rarely occur in isolation. As such they are often perturbed. Perturbations can change the system's state, its dynamics, or both. They can be localized in time, persist over some time, or even permanent. In Fig. (\ref{fig:pertuebations}), we present a few examples: (a) The hitting time of a particle diffusing to a target is altered due to the application of an external field; (b) An unpredictable event (a black swan \cite{taleb2007}) drastically changes the dynamics of financial markets and, inter-alia, the time it takes a stock to reach a designated price; (c) A stochastic search algorithm restarts in attempt to expedite convergence, but only if it failed to cross a checkpoint that indicates proximity to the desired solution.

A fundamental question arises: what is the effect of a perturbation on the mean first-passage time (MFPT)? Specifically, it is not clear if the introduction of a perturbation will increase or decrease the MFPT, and by what magnitude. At face value, it seems that the answer to this question depends on the myriad of details characterizing the underlying first-passage process,the perturbation, and the manner in which the two interact. Yet, we show that only a handful of key observables are important, revealing a universal linear response of the mean first-passage time to perturbations.

Inspired by the profound influence of linear response theory on the understanding of physical systems both in and out of equilibrium, we set forth to establish a linear response theory for first-passage processes. Direct extension of existing results is not possible, since first-passage processes are inherently transient, and thus do not admit a steady-state (equilibrium or not). Consequently, and contrary to the classical theory of linear response, the state of the system right before the perturbation depends on its age. Moreover, the time evolution of first-passage processes may or may not obey Hamiltonian dynamics. Thus, the concept of a perturbation Hamiltonian, pivotal in classic linear response theory, is not well-defined for a general first-passage process.

Clues on how to address the above challenges can be found in the field of stochastic resetting --- a perturbation that has recently attracted considerable interest \cite{Kundu_Preface_2024}. In stochastic resetting, a first-passage process is halted at a random time, and the system is brought back to its initial state. Following the seminal contribution of Evans and Majumdar \cite{Satya_and_Evans_PRL}, a complete framework for first-passage processes under stochastic resetting was developed \cite{FPT_reset, Optimal_stoc_reset, stoc_res_and_app, diffusion_with_opt_reset, sub_unbind, Michaelis_menten_PRE, Sokolov_Renewal}. This revealed that the linear response of the MFPT to the resetting rate is universal and given by $\chi=\frac{\langle T\rangle^2}{2}(1-CV_T^2)$. Here, $\langle T\rangle$ is the unperturbed MFPT, and $CV_T=\sigma(T)/\langle T\rangle$ is the coefficient of variation of the unperturbed first-passage time, where $\sigma(T)$ is the standard deviation of the unperturbed first-passage time. 

Crucially, for $CV_T>1$ we have $\chi<0$, demarcating when stochastic resetting expedites first-passage processes. This CV condition has already found practical use when applying stochastic resetting to molecular dynamics simulations \cite{SR_Enhanced_Sampling}. It was further generalized and adapted to cover cases where resetting is accompanied by a time penalty \cite{Search_with_home_returns}, a branching event \cite{Reset_with_branching}, and to resetting in discrete-time \cite{StocResetDT}.

Here, we elevate from stochastic resetting to the stochastic application of a general perturbation. Similar to stochastic resetting, we circumvent direct extension of classical linear response theory, and consider instead the response of a general first-passage process to a rare --- yet completely arbitrary --- perturbation. We therefore look for an expansion

\begin{equation}\label{eq: lin approx}
    \langle T_\lambda\rangle\simeq\langle T\rangle +\lambda\chi,
\end{equation}

\noindent where $\langle T_\lambda\rangle$ is the MFPT under the perturbation, $\langle T\rangle$ is the underlying MFPT, $\lambda$ is the rate of its application, and $\chi$ is the linear response parameter. We find that the linear response parameter is generally given by

\begin{equation}\label{gen_linear_response}
    \chi = \frac{\langle T\rangle^2}{2}\left(\frac{2\overline{\tau}}{\langle T\rangle}-CV_T^2-1\right),
\end{equation}

\noindent where $CV_T=\sigma(T)/\langle T\rangle$ is once again the coefficient of variation of the unperturbed FPT, and $\overline{\tau}$ is the mean time between the activation of the perturbation and the FPT, conditioned on the perturbation actually occurring. Unlike classic linear response theory, our results continue to hold exactly even for strong perturbations that significantly alter the dynamics.

Demanding that the introduction of a rare perturbation will result in a reduction of the MFPT, i.e. $\chi<0$, we get a generalization of the CV condition that applies for general perturbations. This reads

\begin{equation}\label{gen_CV_cond}
    CV_T^2>\frac{2\overline{\tau}}{\langle T\rangle}-1,
\end{equation}

\noindent which --- similar to classical linear response theory --- highlights the strong dependence of the linear response on fluctuations in the unperturbed FPT. To see this more vividly, observe that a perturbation can reduce the MFPT of a process even if the MFPT after the perturbation, $\overline{\tau}$, is larger than the unperturbed MFPT, $\langle T\rangle$. While this seems counterintuitive, Eq. (\ref{gen_linear_response}) asserts that this will happen whenever fluctuations in the underlying FPT are large enough. The explanation for this effect can be traced back to the inspection paradox \cite{Feller, inspection_paradox}, as we discuss in the next section.

It is easy to verify that the results above hold for the case of resetting. There, in the limit of a vanishingly small resetting rate, the mean time remaining after resetting is simply $\langle T\rangle$, since the probability for multiple resetting events is negligibly small. Substituting the values of $\overline{\tau}$ for resetting with time penalties and resetting with branching also agrees with known results \cite{Search_with_home_returns, Reset_with_branching}. Beyond resetting, the above results hold for arbitrary perturbations and first-passage processes, provided that $\langle T\rangle$, $\sigma(T)$, and  $\overline{\tau}$ are well-defined, i.e., finite.  We stress that the linear response is universal in the sense that it only depends on the above-mentioned moments. Crucially, when these are not known analytically, they can be easily estimated from measured or simulated data.

The remainder of this article is structured as follows. We start by deriving the main result of this work --- Eq. (\ref{gen_linear_response}). Next, we extend this result to cases where the variance of the FPT diverges. We then demonstrate how to use Eq. (\ref{gen_linear_response}) in practice. We do this using two toy models. In both, the underlying process is a 1D drift-diffusion to a target, but it is subject to two different perturbations: resetting with information feedback and the application of a log-potential. While describing these perturbations is straightforward, their effect on the MFPT can hardly be guessed a priori. Yet, we show that Eq. (\ref{gen_linear_response}) can be used to easily map out the linear response phase space. Going beyond linear response, we reveal non-trivial phase transitions of first and second order.

Next, we use our theory to construct an experimental technique that can extract information about the fluctuations in the lifetime of biomolecules using bulk concentration measurements only, i.e., without resorting to single-molecule experiments. To our knowledge, this is the first proposed method to extract higher-order kinetic moments of single molecules from bulk measurements. Specifically, we show that the linear response of the concentration to a perturbation reveals the variance in the molecular lifetime. By going beyond the linear regime, our approach enables the reconstruction of the entire lifetime distribution. We conclude this paper with a discussion and outlook.

\section{Derivation of Eq. (2)}
Consider a first-passage process whose FPT we denote by $T$. We would like to understand the response of the MFPT to an arbitrary, but rare perturbation. To do so, we let the perturbation occur at a random time $P$, which we take from an exponential distribution with rate $\lambda$ that is henceforth considered small in the sense that $\lambda\langle T\rangle\ll1$. The perturbed FPT, $T_\lambda$, is then given by 

\begin{equation}\label{T_lambda definition}
    T_\lambda=
    \begin{cases}
    T & \text{if $T\leq P$}~,\\
    P+\tau(\Vec{q}) & \text{if $T>P$}~.
    \end{cases}    
\end{equation}

\noindent Indeed, if first-passage occurred before the perturbation time $P$ the process completes without interruption. Otherwise, the FPT will occur at $P$ plus the time remaining from the moment of the perturbation till completion. We denote this time by $\tau(\Vec{q})$, and note that it can depend on the state of the system, $\Vec{q}$, at time $P$. Moreover, we allow $\tau(\Vec{q})$ to be deterministic or random. Finally, we will assume that  the statistics of $\tau(\Vec{q})$ is completely determined given $\Vec{q}$, i.e., that it does not depend explicitly on the moment at which the perturbation is applied. We stress that the state $\Vec{q}$ can carry any needed information, e.g., the entire history of the process till the moment the perturbation occurred.

To proceed, we apply the total expectation theorem 
\small
\begin{equation}\label{T_lambda MFPT TET}
\begin{split}
    \langle T_\lambda\rangle &= \Pr(T\leq P)\langle T|T\leq P\rangle+\\
    &+\Pr(P<T)\left(\langle P|P< T\rangle+\langle \tau(\Vec{q})|P< T\rangle\right)=\\
    &=\frac{1-\Tilde{T}(\lambda)}{\lambda}+\lambda\int_{Q}\int_{0}^{\infty}\langle\tau(\Vec{q})\rangle G(\Vec{q},t)e^{-\lambda t} \, dt \, d\Vec{q},
\end{split} 
\end{equation}
\normalsize
\noindent where $\Tilde{T}(\lambda)=\langle e^{-\lambda T}\rangle$ is the Laplace transform of $T$ evaluated at $\lambda$, $G(\Vec{q},t)$ is the probability density to find the process at state $\Vec{q}$ at time $t$, and $\langle\tau(\Vec{q})\rangle$ is the average of $\tau(\Vec{q})$ over all possible realizations. The derivation of Eq. (\ref{T_lambda MFPT TET}) is given in Appendix \ref{appendix: Eq. 5 Derivation}. 

Expanding to first order in the perturbation rate we obtain $\langle T_\lambda\rangle\simeq\langle T\rangle+\chi \lambda,$ with $\chi = \frac{\langle T\rangle^2}{2}\left(\frac{2\overline{\tau}}{\langle T\rangle}-CV_T^2-1\right)$ as defined in Eq. (\ref{gen_linear_response}). For the complete derivation, see Appendix \ref{appendix: dev of main res}. Here, $\overline{\tau}$ is given by
\begin{equation}\label{tau bar definition}    \overline{\tau}=\int_Q\langle\tau(\Vec{q})\rangle\phi(\Vec{q})\,d\Vec{q},
\end{equation}
i.e., it is $\langle\tau(\Vec{q})\rangle$ (the mean of $\tau(\Vec{q})$) averaged over the time averaged propagator

\begin{equation}\label{phi definition}
    \phi(\Vec{q})\equiv\frac{1}{\langle T\rangle}\int_0^\infty G(\Vec{q},t)\, dt,
\end{equation}

\noindent which is the probability measure that governs the system state $\Vec{q}$ at the moment a rare perturbation occurs. Given $\chi$ in Eq. (\ref{gen_linear_response}), it is easy to derive Eq. (\ref{gen_CV_cond}). In our derivation, we assumed that time is continuous. For similar results in discrete time, see Appendix \ref{appendix: discrete time}.

Eq. (\ref{gen_linear_response}) can also be derived by considering the difference between an unperturbed first-passage process and one that has been perturbed at a random time $P$. Noting that the two only differ from the perturbation moment onward, we need only compare the mean completion times starting from this point. For the unperturbed process, this time is given by

\begin{equation}\label{Inspection Paradox}
    \langle T_{\text{res}}\rangle=\frac{\langle T\rangle}{2}\left(1+CV_T^2\right),
\end{equation}

\noindent which is a classical result that follows from the inspection paradox \cite{inspection_paradox, Feller}. For the perturbed process, the remaining time is $\overline{\tau}$, since $\phi(\Vec{q})$ governs the system's state at the time of the perturbation. The difference $\overline{\tau}-\langle T_{\text{res}}\rangle$ captures the mean effect of the perturbation, given that it occurs. Multiplying $\overline{\tau}-\langle T_{\text{res}}\rangle$ by $\lambda\langle T\rangle$, i.e., the probability that the perturbation indeed occurs, yields Eq. (\ref{gen_linear_response}).

Finally, if the FPT of the unperturbed process is taken from a distribution with a power-law tail, $f_T(t)\sim t^{-1-\alpha}$ with $1<\alpha<2$, the unperturbed MFPT is still well-defined and $\phi(\Vec{q})$ is still a proper distribution, but the variance of the unperturbed FPT diverges. Also, in this case, the Laplace transform of the FPT distribution can be written as $\Tilde{T}(\lambda)=1-\langle T\rangle\lambda+c\lambda^\alpha+o(\lambda^\alpha)$, with $c$ being some positive constant. Therefore, for any finite $\overline{\tau}$, Eq. (\ref{T_lambda MFPT TET}) implies that the MFPT at low rates can be approximated as $\langle T_\lambda\rangle\simeq\langle T\rangle-c\lambda^{\alpha-1}$. It is worth emphasizing that for such processes, the leading order in the response is independent of the nature of the perturbation as long as $\overline{\tau}<\infty$. Indeed, the response at low rates is a function of $\alpha$ and $c$ alone, which are properties of the underlying FPT statistics.

\section{Informed Resetting}

We now turn to illustrate the usefulness of our newly established theory via the analysis of two toy models. We start with informed resetting. 

In many settings where stochastic resetting is used to expedite stochastic processes, the decision to reset can be informed by partial knowledge of the system's state—for example, whether it is relatively close to or far from the target. This idea, known as \emph{informed resetting}, avoids unnecessary resets near successful completion and has been shown to lower both MFPTs \cite{Keidar_Blumer2024, Church_Blumer2025} and energetic costs \cite{Tal_Friedman2025} compared to uninformed resetting.

While exact target locations are typically unknown in real search problems, proximity cues are often available—such as structural similarity in molecular dynamics simulations, or environmental gradients in biological foraging. In molecular simulations in particular, the final configuration is usually known, but it is difficult to sample transitions leading to it \cite{bonati2021deep, mendels2018collective}. In such cases, informed resetting protocols can be defined by measurable progress toward the end configuration, offering both theoretical and practical advantages \cite{Keidar_Blumer2024, Church_Blumer2025}.

Despite growing interest in informed resetting, its theoretical underpinnings have remained largely unexplored. The incorporation of feedback complicates analysis, and very little is known about such processes in general. Here, we use our linear response framework to carry out the first analytical treatment of informed resetting in the canonical drift-diffusion model. Our results reveal how information on system state modifies the impact of resetting, allowing one to map the parameter space where such feedback confers an advantage. This example underscores the power of our framework to handle perturbations whose effect is state-dependent, and to yield insight into the tradeoffs between information and control.

\begin{figure}[t]
    \centering
    \includegraphics[width=0.475\textwidth, draft=false]{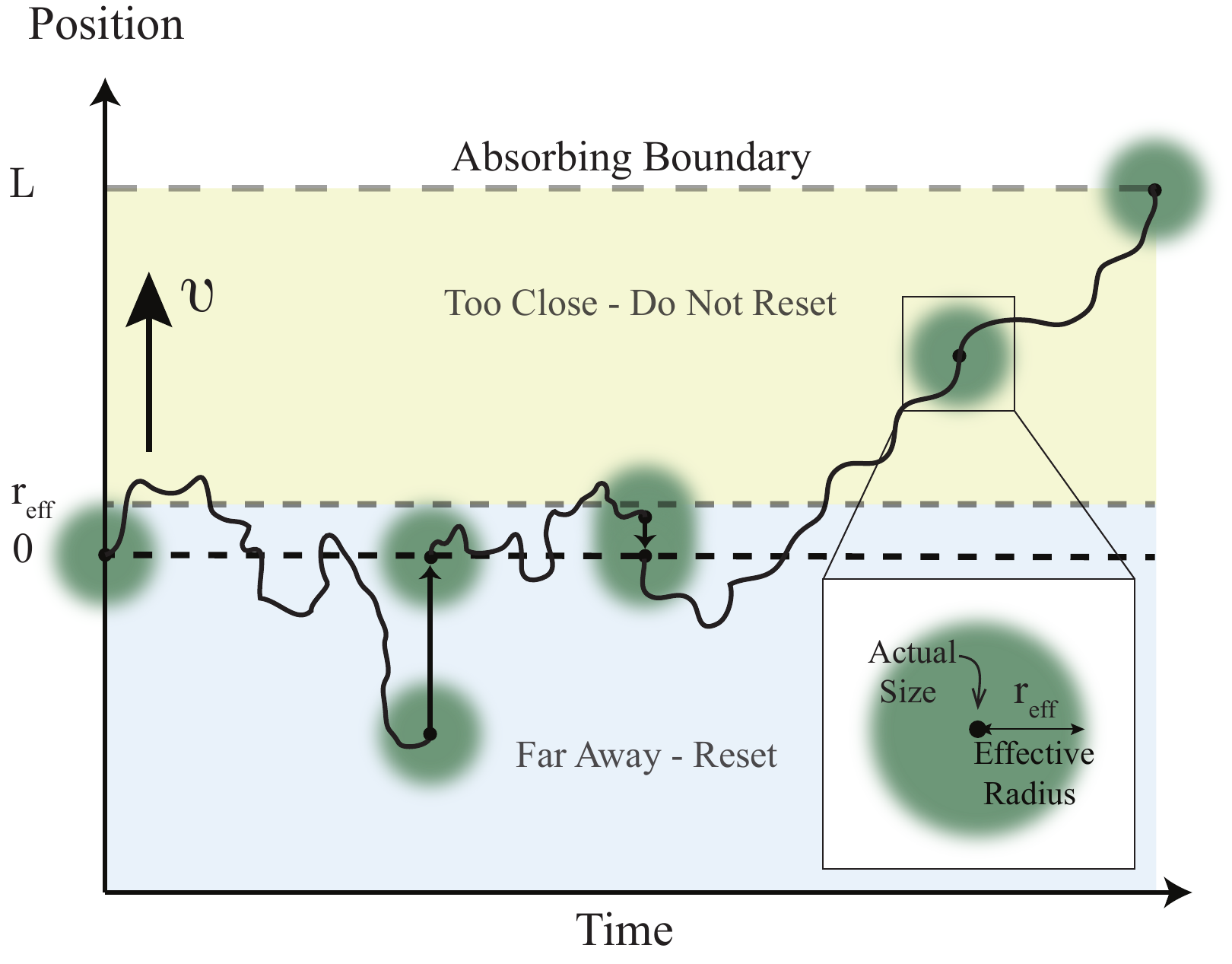}
    \caption{Diffusion under informed resetting. A particle is drift diffusing from the origin to a target located at $L$, with a drift velocity $v$, and a diffusion coefficient $D$. The particle's location is measured at random times, but the measurement has an uncertainty of $r_{\text{eff}}$. Informed resetting is used to outperform ignorant stochastic resetting, by avoiding resetting when this will surely move the particle further away from the target compared to its current position.}
    \label{fig: informed resetting setup}
\end{figure}

Consider the scenario illustrated in Fig. (\ref{fig: informed resetting setup}). A particle is drift diffusing from the origin with a drift velocity $v>0$ and a diffusion coefficient $D$. At $L>0$, there is an absorbing boundary, and we are interested in the mean absorption time. The particle's actual size is negligible, but due to the limitations of the experimental apparatus, it cannot be localized precisely and is seen as a smudge of radius $r_{\text{eff}}$. At some rate $\lambda$, we measure the particle's location. If we are sure that the particle is closer to $L$ than its initial position, i.e., it is in position $x>r_{\text{eff}}$, we do nothing. Otherwise, we bring the particle back to the origin and reset its motion. We call this perturbation: `informed resetting'.

\begin{figure*}[t]
    \centering
    \includegraphics[width=\textwidth]{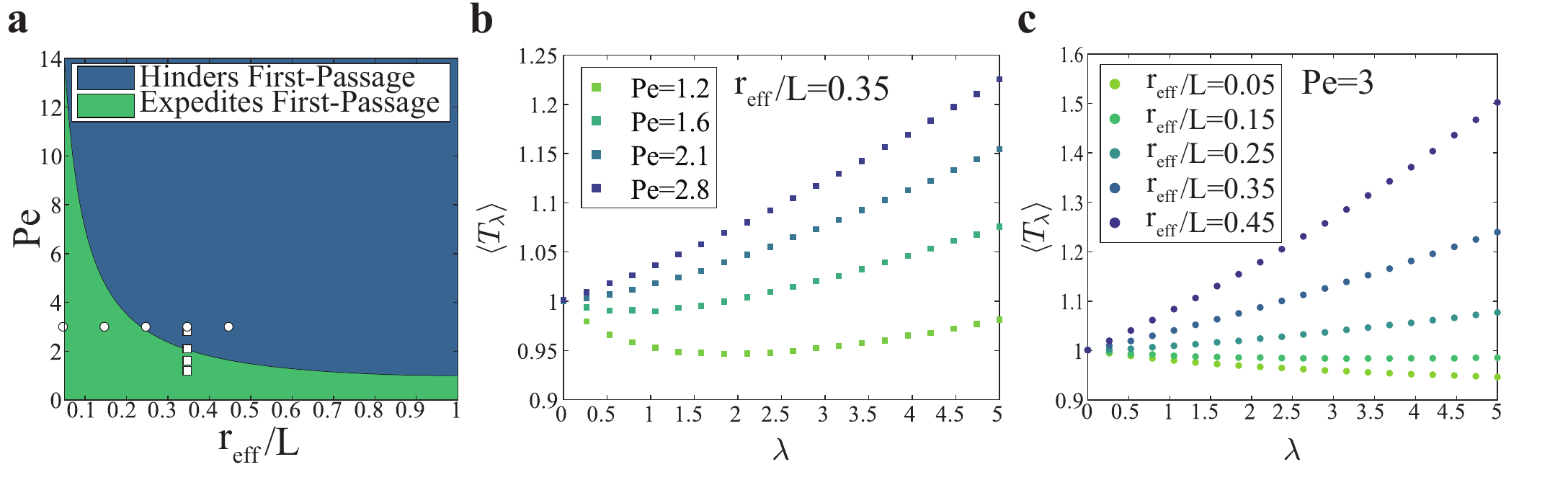}
    \caption{\textbf{a}, The two phases of drift-diffusion with informed resetting are constructed using Eq. (\ref{linear response informed}). When the sign of the response is negative, informed resetting expedites first-passage (green phase). The converse happens when the response is positive (blue phase). \textbf{b}, The MFPT vs. the resetting rate $\lambda$. Squares come from Langevin simulations with $r_{\text{eff}}/L=0.35$ and different $Pe$. Simulations agree with the prediction for a transition from a positive to a negative linear response at $Pe\simeq 2.05$. \textbf{c}, The MFPT vs. the resetting rate $\lambda$. Circles come from Langevin simulations with $Pe=3$ and different $r_{\text{eff}}/L$. Simulations agree with the prediction for a transition from a positive to a negative linear response at $r_{\text{eff}}/L\simeq0.24$. All simulations were conducted with $L=v=1$.} 
    \label{fig: informed resetting graphs}
\end{figure*}

We would like to understand the effect of informed resetting on the MFPT of drift-diffusion. Two limits of this problem have already been considered and solved. When $r_{\text{eff}}=L$, resetting is ignorant of the particle's position, and the problem boils down to regular stochastic resetting of drift-diffusion. This problem was solved in \cite{RayPeclet}, where the authors calculated the MFPT as a function of the resetting rate $\lambda$ from which one can read the linear response, $\chi$, as $\chi=\frac{L^2}{2v^2}\left[1-\frac{1}{Pe}\right]$. Here, $Pe\equiv\frac{Lv}{2D}$ is the P\'eclet number, and note that $\chi$ changes sign at $Pe=1$, which corresponds to $CV_T=1$ \cite{RayPeclet}. 

In the other extreme, $r_{\text{eff}}=0$, we have full information on the particle's position. The MFPT can then be obtained by mapping our problem onto the problem of asymmetric stochastic resetting that was presented in \cite{Asymmetric_SR}. Using tools developed there, we find  $\chi=\frac{L^2}{4v^2Pe^2}\left[e^{-2Pe}-1\right]$  (Appendix \ref{appendix: full information}), which is always negative since resetting with full information is guaranteed to expedite first-passage. 

The linear response function in the limits of zero and full information was calculated based on an explicit expression for the MFPT as a function of the perturbation rate. Yet, finding the MFPT at an arbitrary perturbation rate is not always feasible, and commonly not a simple task. Indeed, even for the relatively simple toy model considered herein, an exact solution for the MFPT at intermediate values of $r_{\text{eff}}$ cannot be obtained using the existing stochastic resetting framework \cite{stoc_res_and_app}, and requires particular treatment. Next, we circumvent brute force solution of the problem and show that the linear response of drift-diffusion to informed resetting can be found with little effort using Eq. (\ref{gen_linear_response}). 

To find the linear response given in Eq. (\ref{gen_linear_response}), we need the MFPT and $CV_T$ of the underlying process, and  $\overline{\tau}$ which is given by Eq. (\ref{tau bar definition}). For drift-diffusion, we have $\langle T\rangle=L/v$ and $CV_T^2=Pe^{-1}$ \cite{redner_book}. To find $\overline{\tau}$ we first need to find $\langle\tau(x)\rangle$, which is the mean time to a first-passage from the moment of the perturbation, given that this was activated when the particle was at $x$. 

Since we are interested in the linear response, it is enough to consider the limit of a rare perturbation ($\lambda\to 0$). In this limit, one can safely neglect multiple occurrences of the perturbation, and we find

\begin{equation}\label{tau_x informed}
    \langle\tau(x)\rangle=\begin{cases}
        \frac{L-x}{v}& \text{if $x>r_{\text{eff}}$}~,\\
        \frac{L}{v}&\text{if $x\leq r_{\text{eff}}$~.}
    \end{cases}
\end{equation}

\noindent To understand Eq. (\ref{tau_x informed}), observe that the upper branch accounts for the case where the perturbation catches the particle at $x>r_{\text{eff}}$, and there is no resetting.  The particle then continues to drift-diffuse, reaching the target after $(L-x)/v$ units of time on average. The lower branch accounts for the case where the perturbation catches the particle at $x\leq r_{\text{eff}}$. The particle's location is reset to the origin, from which it will take $L/v$ units of time on average to reach the target. 

Next, we need to find the mean of $\langle\tau(x)\rangle$ using the probability measure $\phi(x)$ defined in Eq. (\ref{phi definition}). Plugging in the known propagator for drift-diffusion with an absorbing boundary $G(x,t)=\frac{1}{\sqrt{4\pi Dt}}e^{-\frac{(x-vt)^2}{4Dt}}\left(1-e^{-\frac{L(L-x)}{Dt}}\right)$ \cite{redner_book}, we find
\begin{equation}\label{phi drift diffusion}
     \phi(x)=\frac{1}{L}\left(e^{\frac{v}{2D}(x-|x|)}-e^{-2Pe}e^{\frac{vx}{D}}\right).
\end{equation}

\noindent Integrating the product of Eqs. (\ref{tau_x informed}) and (\ref{phi drift diffusion}) results in $\overline{\tau}$. Inserting into Eq. (\ref{gen_linear_response}) gives the following linear response

\begin{equation}\label{linear response informed}
\begin{split}
    \chi=\frac{L^2}{2v^2}&\left[\left(\frac{r_{\text{eff}}}{L}\right)^2-\frac{1}{2Pe^2}+\right.\\
    &+\left.\frac{e^{-2Pe\left(1-\frac{r_{\text{eff}}}{L}\right)}}{Pe}\left(\frac{1}{2Pe}-\frac{r_{\text{eff}}}{L}\right)\right]~.
\end{split}  
\end{equation}

\noindent Plugging $r_{\text{eff}}/L=1$, gives back the linear response of stochastic resetting of a drift-diffusion process \cite{RayPeclet}, and plugging $r_{\text{eff}}/L=0$ we get the same result obtained in \cite{Asymmetric_SR}, as explained in the Appendix \ref{appendix: full information}.

\begin{figure*}[t!]
    \centering
    \includegraphics[width=0.8\textwidth]{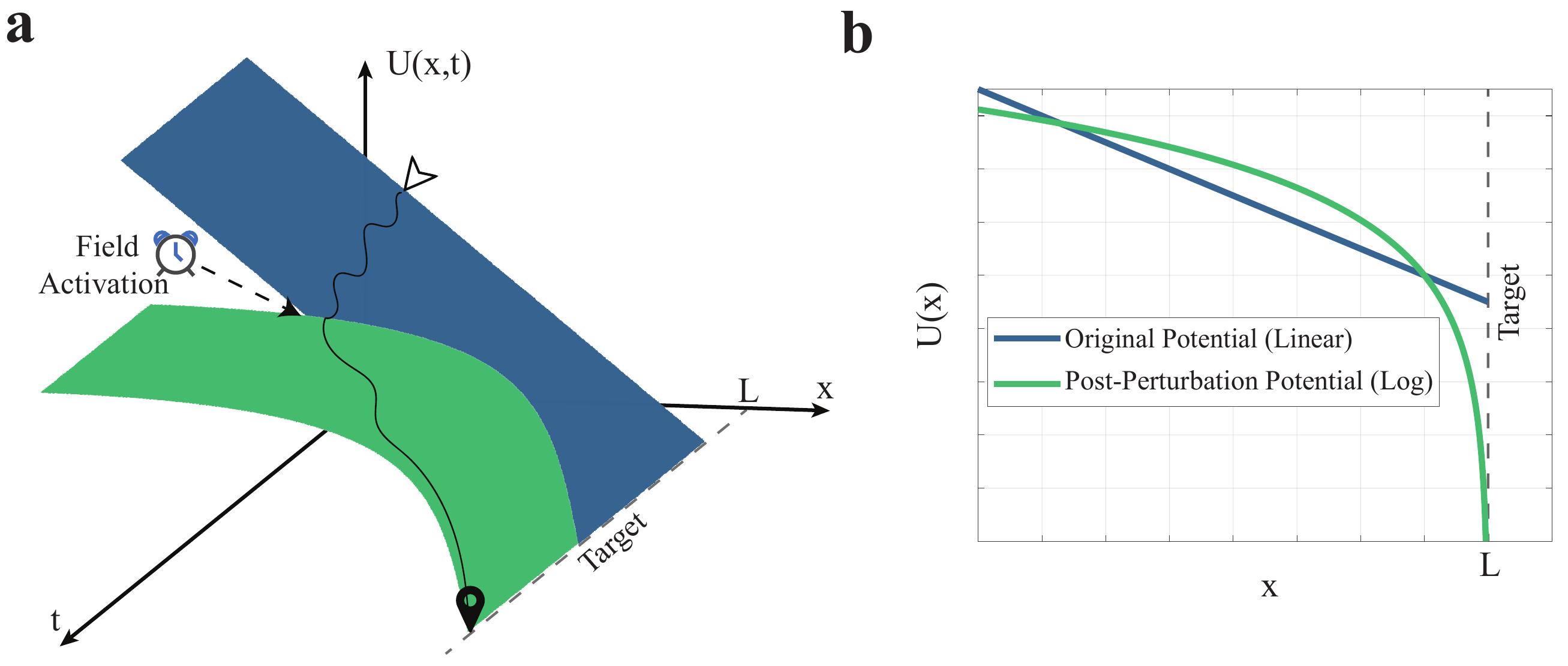}
    \caption{\textbf{a}, An illustration of the problem of stochastic field activation. A particle starts drift-diffusing from the origin to a target located at $L$. At some random point in time, a field is activated and the particle continues its diffusion on a logarithmic potential. \textbf{b}, The potential energy before and after field activation. Before the field is activated the potential energy is linear and given by $U_{lin}(x)=-F_0x$. After the field is activated the potential energy is logarithmic and given by $U_{log}(x)=U_0\ln{\left(L-x\right)}$.
    }
    \label{fig: lin-log demo}
\end{figure*}

Taking the sign of the linear response, we create a phase space separating cases where informed resetting expedites first-passage from cases where it has the opposite effect. This phase space is shown in Fig. (\ref{fig: informed resetting graphs}a). Going from the no information case ($r_{\text{eff}}=L$) to the full information case ($r_{\text{eff}}=0$), we see that the class of drift-diffusion processes that can be expedited by informed resetting becomes larger. Namely, more information allows acceleration at higher P\'eclet numbers.

To corroborate the separatrix predicted by Eq. (\ref{linear response informed}), we performed Langevin simulations of drift-diffusion with informed resetting. First for $r_{\text{eff}}/L=0.35$ and different $Pe$ [squares in Fig. (\ref{fig: informed resetting graphs}a) and Fig. (\ref{fig: informed resetting graphs}b)], and then for $Pe=3$ and different values of $r_{\text{eff}}/L$ [circles in Fig (\ref{fig: informed resetting graphs}a) and Fig. (\ref{fig: informed resetting graphs}c)]. In both cases, the transition from acceleratory to inhibitory response agrees with the one predicted by Eq. (\ref{linear response informed}).

How accurate should a position measurement be in order to expedite the first-passage of drift-diffusion using informed resetting? Seemingly, answering this question requires a detailed solution to the particular problem illustrated in Fig. (\ref{fig: informed resetting setup}). Yet, we showed that the latter can be skipped altogether. We applied Eq. (\ref{gen_linear_response}), which allows us to answer the posed question directly. Eq. (\ref{gen_linear_response}) can thus be used as a quick and simple tool to understand the response of a first-passage process to a perturbation while avoiding the difficulties of solving the FPT problem explicitly.

\section{Stochastic Activation of a Field}

We will now show that the presented theory can be used to unravel non-trivial phenomena that emerge in a seemingly simple setup where a field is activated randomly in time. We again consider the example of a particle drift diffusing to a target at $L$. The particle starts at the origin, with a drift velocity $v>0$, and a diffusion coefficient $D$. The diffusion coefficient follows the Einstein-Smoluchowski relation: $D=(\beta\zeta)^{-1}$ where $\zeta$ is the drag coefficient and $\beta$ is the inverse temperature. The potential describing this problem is $U_{lin}(x)=-F_0x$, where $F_0=v\zeta$. We would like to understand the response of the MFPT to a field, which is activated at a random time and kept `on' until first-passage occurs. For concreteness and analytical tractability, we consider a post-perturbation potential that is logarithmic $U_{log}(x)=U_0\ln{(L-x)}$, with $U_0>0$. An illustration of the process is presented in Fig. (\ref{fig: lin-log demo}). 

Observe that both the magnitude and sign of the response to the considered perturbation are not trivial. This is because they depend on the \textit{random} position of the particle at the moment of field activation. Indeed, if the particle is near the target, where the logarithmic potential is much steeper than the linear counterpart, it will feel a strong pull that would shorten its MFPT. On the other hand, if the particle is far away from the target, its MFPT will increase since the logarithmic potential is almost flat there (compared to the linear). 

To proceed, we first analyze the linear response to the perturbation. To this end, we need to compute the terms that appear on the right-hand side of Eq. (\ref{gen_linear_response}). Because drift-diffusion is the underlying process prior to the perturbation, we have $\langle T\rangle=L/v$ and $CV_T^2=Pe^{-1}$ as in the previous section \cite{redner_book}. To obtain $\overline{\tau}$, we first find $\langle\tau(x)\rangle$: the mean time remaining to first-passage given that the particle was at $x$ at the moment the perturbation occurred. By construction, this is nothing but the MFPT of a particle diffusing from $x$ to $L$ on a logarithmic potential \cite{Bray_XY_Model,RayLog}
\begin{equation}\label{log MFPT}
    \langle\tau(x)\rangle=\begin{cases}
        \frac{(L-x)^2}{2D(\beta U_0-1)} & \text{if $\beta U_0>1$}~,\\
        \infty & \text{if $\beta U_0<1$}~.
    \end{cases}
\end{equation}
\noindent To get $\overline{\tau}$, we need to average Eq. (\ref{log MFPT}) with respect to the appropriate position distribution $\phi(x)$. For a drift-diffusion process, this is given by Eq. (\ref{phi drift diffusion}) as explained in the previous section.

Following the steps above, we substitute $\langle T\rangle$, $CV_T^2$, and $\overline{\tau}$ into the right hand side of Eq. (\ref{gen_linear_response}). The resulting linear response diverges for $\beta U_0<1$, which is easy to see from Eq. (\ref{log MFPT}). For $\beta U_0>1$ we find 

\begin{equation}\label{linear response change of potential}
    \chi=\frac{2DL}{3v^3}\frac{Pe^2+\frac{3}{2}(2-\beta U_0)(Pe+1)}{\beta U_0-1}.
\end{equation}

\noindent From here it is clear that the response is finite and positive whenever $1<\beta U_0\leq2$. Namely, in this regime stochastic activation of a logarithmic field at low rates will always hinder first-passage. 

\begin{figure}[t]
    \centering
    \includegraphics[width=0.9\linewidth]{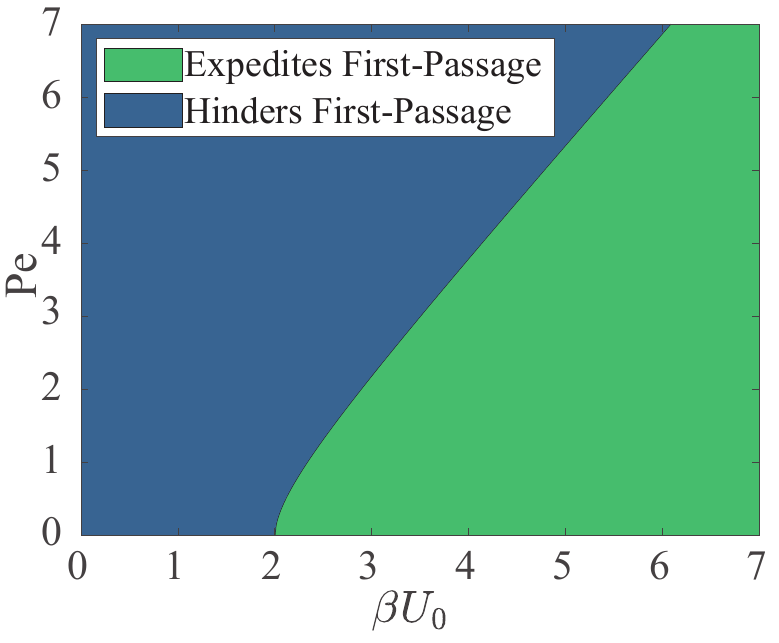}
    \caption{The phase space for the problem of stochastic field activation that is illustrated in Fig. (\ref{fig: lin-log demo}).}
    \label{fig: log first phase space}
\end{figure}

When $\beta U_0>2$, the sign of the response depends on $Pe$. Namely, the response is positive when $Pe>Pe^\star$ with $Pe^\star=\frac{3(\beta U_0-2)}{4}\left(1+\sqrt{(3\beta U_0+2)/(3\beta U_0-6)}\right)$, and negative otherwise. These critical $Pe$ separate the part of the phase space where field activation hinders first-passage from the part where field activation expedites first-passage. The phase space, constructed using Eq. (\ref{linear response change of potential}), is given in Fig. (\ref{fig: log first phase space}).

\subsection{Beyond Linear Response}
\begin{figure*}[t!]
    \centering
    \includegraphics[width=0.9\textwidth]{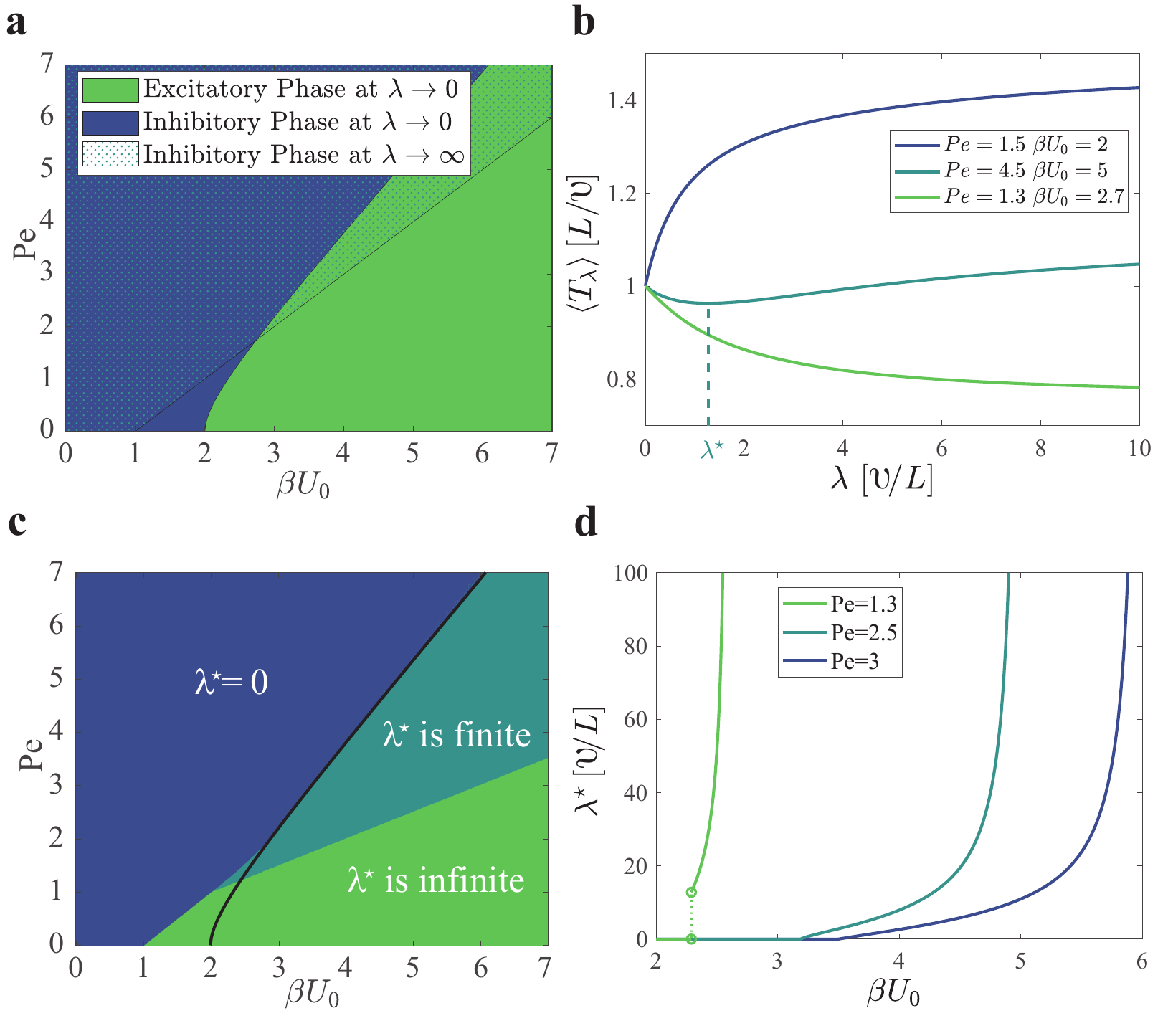}
    \caption{\textbf{a}, The region of phase space where immediate ($\lambda\to\infty$) activation of a logarithmic potential hinders first-passage (dotted), superimposed on the linear response phase space ($\lambda\to0$) from Fig. (\ref{fig: log first phase space}). \textbf{b}, The MFPT as a function of the perturbation rate $\lambda$ for three different pairs of $Pe$ and $\beta U_0$. Three behaviors of the optimal rate $\lambda^\star$ can be seen, it is either zero, finite, or infinite. \textbf{c}, The phase space for the optimal switching rate. A black line indicating the separatrix at $\lambda\to0$, that was obtained using linear response analysis. \textbf{d}, Plots of the optimal rate $\lambda^\star$ that minimizes the MFPT as a function of $\beta U_0$ for $Pe=1.3$ and $Pe=2.5$. It can be seen that the transition from $\lambda^\star=0$ to the finite $\lambda^\star$ phase can either be continuous (for $Pe=2.5$) or discontinuous (for $Pe=1.3$). Dashed lines at $\beta U_0=2Pe$ indicate the predicted transition to the infinite $\lambda^\star$ phase.} 
    \label{fig: lin-log phase space}
\end{figure*}

So far, we have focused on the linear response to a rare perturbation ($\lambda\to 0$). Yet, in many cases, the response to an immediate application of a perturbation ($\lambda\to\infty$) can also be obtained with little effort, allowing analysis and insight beyond linear response.

In the above example, the behavior at very large perturbation rates can be obtained by noting that the particle has no time to move before the field is activated. In addition, because the field is activated once and for all, the MFPT in this limit is given by setting $x=0$ in Eq. (\ref{log MFPT}). To create the phase space at $\lambda\to\infty$, we compare this MFPT  to $L/v$ (MFPT on linear potential) and determine which one is greater. In Fig. (\ref{fig: lin-log phase space}a), we take the $\lambda\to0$ phase space from Fig. (\ref{fig: log first phase space}) (colored) and plot on top of it the region in phase space where immediate ($\lambda\to\infty$) activation of a logarithmic potential hinders first-passage (dotted). It then becomes clear that a wedge-shaped overlap exists between the region in phase space where rare field activation expedites first-passage (green), and the region in phase space where immediate field activation hinders first-passage.  In this wedge-shaped region, the MFPT  decreases with $\lambda$ when it is small, but its value at very high perturbation rates is larger than its value at $\lambda=0$. It follows that there must be a finite perturbation rate, $\lambda^\star$, that minimizes the MFPT.

This observation raises the suspicion that the phase space of this $\lambda^\star$, i.e. the frequency of field activation that minimizes the MFPT, can be divided into regions where $\lambda^\star$ is either finite, zero, or infinite. The wedge-shaped region must lay where $\lambda^\star$ is finite. Yet, the optimal rate itself cannot be obtained from linear response, but only via an expression that gives the MFPT at an arbitrary field activation rate. We find this expression by use of Eq. (\ref{T_lambda MFPT TET}), which is valid for arbitrary values of $\lambda$.

Perturbing a process repeatedly affects $\langle\tau(\Vec{q})\rangle$ in Eq. (\ref{T_lambda MFPT TET}), which is another way of saying that this quantity generally depends on the perturbation rate. A notable example is stochastic resetting, where the mean residual time after resetting can be approximated by $\langle T\rangle$ only when the probability of having multiple resetting events is negligible. The problem considered here is different since the field is activated once and for all. It follows that $\langle\tau(\Vec{q})\rangle$ is given by Eq. (\ref{log MFPT}) regardless of the perturbation rate $\lambda$. Integrating over time in Eq. (\ref{T_lambda MFPT TET}), we find $\langle T_\lambda\rangle=\frac{1-\Tilde{T}(\lambda)}{\lambda}+\lambda\int_Q\langle\tau(\Vec{q})\rangle\Tilde{G}(\Vec{q},\lambda)\, d\Vec{q}$, where $\Tilde{T}(\lambda)$ and $\Tilde{G}(\Vec{q},\lambda)$ are the Laplace transforms of the FPT distribution and the propagator of the unperturbed process evaluated at $\lambda$, respectively.

A closed-form solution for the MFPT at an arbitrary perturbation rate is presented in the Appendix \ref{appendix: field activation}. From it, the MFPT for any pair of $Pe$ and $\beta U_0$ can be found as a function of $\lambda$, and consequently, the corresponding $\lambda^\star$. In Fig. (\ref{fig: lin-log phase space}b), we plot the MFPT as a function of $\lambda$ for three different pairs of $Pe$ and $\beta U_0$, covering scenarios where $\lambda^\star=0,\infty$, or is some finite value, as indicated on the plot. 

The complete phase space for the optimal rate is presented in Fig. (\ref{fig: lin-log phase space}c) with a solid black line indicating the separatrix at $\lambda\to0$. We note that for large values of $Pe$ and $\beta U_0$ the separatrix at $\lambda\to0$ predicts correctly the transition between the $\lambda^\star=0$ and the finite $\lambda^\star$ phases. The transition between the finite $\lambda^\star$ phase and the $\lambda^\star=\infty$ phase can be obtained analytically. In Appendix \ref{appendix: field activation}, we show that it occurs at $\beta U_0=2Pe$, and that near this transition $\lambda^\star\sim(2Pe-\beta U_0)^{-1}$.

Fig. (\ref{fig: lin-log phase space}d) illustrates how $\lambda^\star$ varies with $\beta U_0$ for two different values of $Pe$: $Pe=1.3$ (blue) and $Pe=2.5$ (orange). The dashed lines at $\beta U_0=2Pe$ mark the analytically predicted transition between the finite and infinite $\lambda^\star$ phases. The transition between the finite $\lambda^\star$ and $\lambda^\star=0$ phases can be either first or second order: a discontinuous transition is found for $Pe=1.3$ and a continuous one for $Pe=2.5$.

The emergence of both first- and second-order phase transitions in a problem as simple as the application of a field is non-trivial and surprising. More than anything, it emphasizes how little we know about the response of first-passage times to perturbations. Extending this analysis to other systems is expected to shed more light on the rich landscape of perturbed first-passage processes.

\section{Molecular Fluctuations from Bulk Measurements}

Up until now, we have shown how our theory can be used to predict the effect of perturbations on FPT processes. We now show that it can also be used in the reverse direction—to extract information about the underlying FPT process itself by analyzing how it responds to perturbations. This idea opens the door to a powerful inference strategy: using externally applied perturbations to uncover hidden microscopic fluctuations.

It is widely believed that fluctuations at the single-molecule level are inaccessible through bulk measurements. This is because the latter report on observables that are averaged over a huge number of molecules. However, the results we have derived reveal that---analogously to classical equilibrium systems---fluctuations in first-passage times govern the response of the mean to perturbations. Leveraging this insight, we propose an experimental approach that enables the inference of single-molecule lifetime fluctuations solely from bulk concentration measurements. This new approach offers an accessible alternative to single-molecule techniques.

\begin{figure*}[t!]
    \centering    \includegraphics[width=0.9\textwidth]{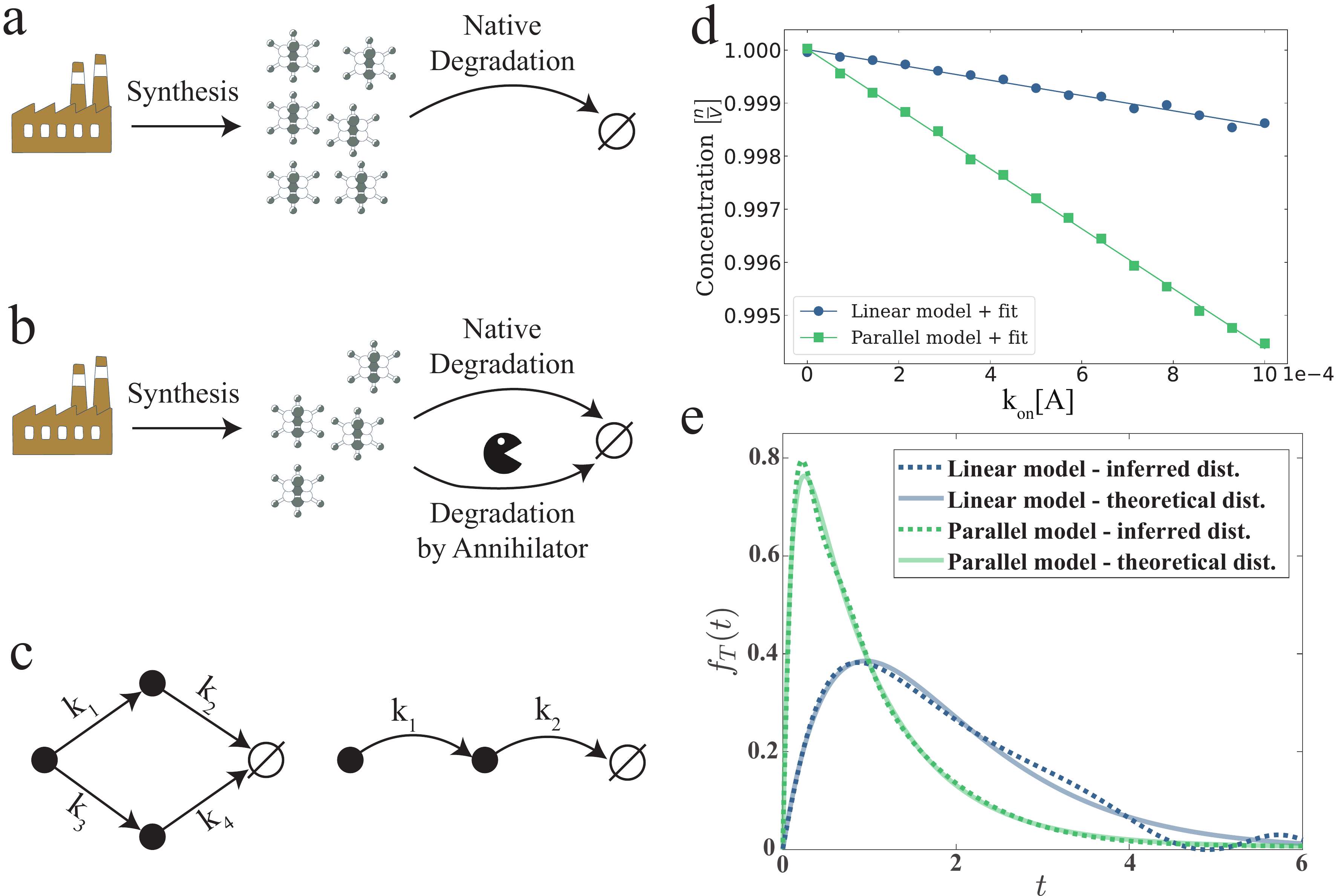}
    \caption{\textbf{a,} Molecules are synthesized and naturally degraded. \textbf{b,} The system is perturbed by the introduction of an annihilator, which degrades the molecules by a different pathway. \textbf{c,} Native degradation may occur by various mechanisms that cannot be differentiated based on the steady-state concentration. Here we consider two possible degradation mechanisms: linear and parallel. The $k_i$'s stand for transition rates. \textbf{d,} Concentration of the molecule of interest as a function of the annihilator concentration. The two models in panel c, yield the same concentration without the annihilator, but their response to its introduction is different. Symbols come from simulation, and solid lines are fits to Eq. (\ref{eq: concentration vs. annihilator}). Error bars are smaller than the symbols. \textbf{e,} Inferred lifetime distribution from bulk measurements compared to the ground truth for the two models depicted in panel \textbf{c}. Simulation details are given in Appendix \ref{appendix: simulation details}.}
    \label{fig: little law}
\end{figure*}

Consider a compound synthesized within a cell, such as a protein or RNA molecule \cite{Elgart2010, Hilfinger2016, Szavits-Nossan2024}. Each molecule has a finite lifetime and is eventually degraded (Fig. \ref{fig: little law}a). According to Little’s law \cite{Little_1961}, and assuming ergodicity in the number of molecules $N$ per cell, the mean number of molecules is given by
\begin{equation}\label{eq: Little law}
    \langle N\rangle=\frac{\langle T\rangle}{\langle S\rangle},
\end{equation}
where $T$ is the lifetime of the molecule and $S$ is the time between consecutive synthesis events. 

In bulk measurements, the concentration is $n\langle N \rangle/V$, where $n$ is the number of cells and $V$ is the sample volume. As a result, lifetime fluctuations are masked and cannot be inferred directly from concentrations. However, a molecule’s lifetime can be regarded as an FPT---starting with synthesis and ending with degradation---whose mean corresponds to the MFPT. By introducing a rare perturbation, we can alter the mean lifetime, with the shift quantified by Eq. (\ref{gen_linear_response}). This, in turn, modifies the bulk concentration via Eq. (\ref{eq: Little law}), opening a peephole into single-molecule fluctuations. 

By selecting a perturbation with a known $\overline{\tau}$, one can infer the value of $CV_T^2$ from the concentration’s response to the perturbation rate. This value provides key insights into the underlying degradation mechanism \cite{Moffitt2014}. For instance, its inverse, $1/CV_T^2$, serves as a lower bound on the number of steps involved in a first-passage process \cite{Aldous_Shepp1987, Moffitt2010PNAS, Moffitt2014}. This principle has been applied to show that the molecular motor kinesin hydrolyzes a single ATP molecule per step \cite{Schnitzer1997}, and to elucidate the mechanism of the DNA packaging motor in Bacteriophage $\varphi 29$ \cite{Moffitt2010PNAS}. Furthermore, if $CV_T^2 > 1$, the degradation process cannot be described by a linear, nearest-neighbor hopping mechanism and must involve branching pathways \cite{Keilson1979, Moffitt2014, Satija2020}.

As a perturbation, we will consider the introduction of an annihilator that degrades the molecule of interest upon contact (diffusion-limited annihilation). The annihilator perturbs the degradation process at a constant rate $\lambda = k_{\text{on}}[A]$, where $[A]$ is the concentration of the annihilator and $k_{\text{on}}$ is its binding affinity (Fig. \ref{fig: little law}b). The concentration of the molecule of interest can then be measured as a function of $[A]$. Using Eqs. (\ref{eq: Little law}) and (\ref{gen_linear_response}), the resulting concentration profile can be calculated. 

Since annihilation is diffusion-limited, the mean residual time following the perturbation, i.e., after contact with the annihilator, is negligible ($\overline{\tau}\simeq0$). The perturbed concentration is thus given by
\begin{equation}\label{eq: concentration vs. annihilator}
\begin{split}
    C([A])=&\frac{n\langle T_{k_{on}[A]}\rangle}{V\langle S\rangle}\simeq C(0)\\
    &-\frac{k_{on}V\langle S\rangle\left[C(0)\right]^2}{2n}\left(CV_T^2+1\right)[A],
\end{split}
\end{equation}
where $C(0)$ is the unperturbed concentration. Notably, the relative fluctuations in the molecule's lifetime ($CV_T^2$) directly influence the response. Thus, by fitting a straight line to the measured concentration curve, one can extract hidden information: the variance in the molecule's lifetime. This, in turn, can help validate or rule out proposed degradation mechanisms.

To illustrate our approach, we consider two Markov models for the degradation process: a linear model and a parallel model (Fig. \ref{fig: little law}c, and details in Appendix \ref{appendix: simulation details}). Notably, both mechanisms can yield the same steady-state concentration in the absence of the annihilator. We simulated each system under varying perturbation rates, corresponding to different annihilator concentrations (Fig. \ref{fig: little law}d). Full simulation details are presented in Appendix \ref{appendix: simulation details}. In both cases, the resulting concentration curves are well approximated by linear fits over the simulated range. Although the models produce identical concentrations without perturbation, their responses to the annihilator differ significantly. Using Eq. (\ref{eq: concentration vs. annihilator}), and assuming all other parameters are known, we inferred $CV_T^2$ from the linear response. For both models, the theoretical values lie within the error bounds of the inferred parameters, and the relative error in estimating $CV_T^2$ was $\approx10^{-2}$.

By extracting the value of $CV_T^2$ from simulations of the two models, we gain access to key mechanistic insights. For the linear model, we obtained an estimated value of $CV_T^2=0.510\pm0.052$, closely matching the true value of $0.5$. This suggests the presence of at least $1.96\pm0.20$ steps in the degradation process \cite{Moffitt2010PNAS, Moffitt2014,Aldous_Shepp1987}. This is consistent with the two-stage nature of the degradation process, as illustrated in Fig. (\ref{fig: little law}c). For the parallel model, we obtained $CV_T^2=4.912\pm0.051$, again in close agreement with the true value of $4.946$. This value exceeds one, ruling out a simple linear nearest-neighbor hopping process as an adequate description of the degradation mechanism \cite{Keilson1979, Moffitt2014, Satija2020}. Notably, all of this information was previously inaccessible through bulk concentration measurements alone. However, by measuring the response of the steady-state concentration, our approach now provides a compelling bulk-level alternative to single-molecule experiments.

We now demonstrate how our theory can also be used to infer \textit{the entire lifetime distribution} from bulk measurements. Returning to Eq. (\ref{T_lambda MFPT TET}), we see that $\langle\tau(\Vec{q})\rangle=0$ for the scenario considered herein. Therefore, using Eq. (\ref{eq: Little law}), we find that the concentration is given by $C([A])=\frac{n\langle T_{k_{\text{on}}[A]}\rangle}{V\langle S\rangle}=\frac{n}{V\langle S\rangle}\frac{1-\Tilde{T}(k_{\text{on}}[A])}{k_{\text{on}}[A]}$, and solving for the Laplace trasform of the lifetime gives
\begin{equation}
\begin{split}
    \Tilde{T}(k_{\text{on}}[A])=1-\frac{V\langle S\rangle k_{\text{on}}}{n}[A] C([A]).
\end{split}
\end{equation} 

This relation shows that by measuring the concentration of the molecule of interest, $C([A])$, as a function of the annihilator concentration $[A]$, one directly obtains the Laplace transform of the lifetime distribution. Subsequently, by numerically inverting this transform, one can reconstruct the full probability density function of the molecular lifetime. 

In Fig. (\ref{fig: little law}e), we show simulation results for the proposed experiment for the two models presented in Fig. (\ref{fig: little law}c). A good agreement can be seen between the inferred distributions and the ground truth. Full simulation details are given in Appendix \ref{appendix: simulation details}. This is a powerful demonstration of the utility of our theory. Previously, accessing such information would require measuring the lifetime of a large number of single molecules, therefore requiring special equipment and expertise. In contrast, we only used thirty bulk concentration measurements (one at each annihilator concentration) for each model. Moreover, we did it without assuming anything about the underlying degradation process, making this method extremely robust.

\section{Summary and Outlook}

In this article, we studied the linear response of the MFPT to a rare, yet arbitrary, perturbation. Surprisingly, we found a fairly simple expression---Eq. (\ref{gen_linear_response})---that captures the linear response of any FPT process (with finite first and second moments) to an arbitrary perturbation. The resulting expression depends on the variance in the FPT of the unperturbed process, echoing a striking similarity to well-known results from classical linear response theory \cite{FDR_review}.

We also dealt with cases where the second moment of the FPT diverges but the MFPT is finite. In those cases, the response to the perturbation is not linear. More importantly, it is a function of the underlying process only, and is independent of the nature of the perturbation itself.

The universal linear response relation for the MFPT significantly streamlines the analysis of scenarios that previously required ad-hoc treatment. For example, we used it to quantify how the addition of information feedback helps stochastic resetting expedite drift-diffusion processes. We then went on to show that our results can also help determine the linear response to the activation of a field. Specifically, we analyzed the effect of changing the potential in a diffusion process from linear to logarithmic. This analysis hinted that the optimal perturbation rate, which minimizes the MFPT, exhibits rich behavior. Extending the analysis beyond the linear regime confirmed the presence of non-trivial first- and second-order phase transitions.

We further demonstrated a practical application of our theory: inferring single-molecule lifetime fluctuations from bulk concentration measurements. Contrary to conventional wisdom, which holds that such fluctuations are masked in bulk due to averaging over large populations, we showed that the response of the concentration to a rare perturbation encodes the variance of the underlying lifetime distribution. Specifically, we showed that the concentration's linear response to the introduction of a diffusion-limited annihilator reveals the coefficient of variation in the molecule's lifetime---a key indicator of the underlying degradation mechanism. We demonstrated how this could e.g., be used to distinguish between linear and parallel degradation pathways, which otherwise yield indistinguishable steady-state concentrations.

We went further and demonstrated that our approach can be extended beyond linear response. By measuring the concentration as a function of annihilator concentration, one can reconstruct the Laplace transform of the lifetime distribution. Numerical inversion of this transform yields the full lifetime probability density function, enabling access not just to the variance but also to higher-order statistics. To our knowledge, this is the first method capable of extracting kinetic fluctuations—and even the full single-molecule lifetime distribution—directly from bulk data. It bypasses the need for single-molecule experiments, offering a significantly more accessible route to mechanistic insights in molecular systems.

The examples we considered illustrate two powerful use cases of our framework: (i) predicting and explaining how a first-passage process responds to perturbations, and (ii) inferring hidden properties of an unperturbed system by measuring its response. For the former task, we have recently shown that our theory can be used to optimize perturbations for improved training of machine learning models \cite{Meir_2025}.

The only information needed to determine the linear response of a first-passage process to a perturbation is the mean and standard deviation of the unperturbed FPT, as well as the mean residual time to first-passage after the activation of the perturbation. In some cases, this information can be acquired using known analytical formulas. More importantly, when these are lacking, the required moments can be obtained from simulated or measured data which are accessible even in the absence of full knowledge of the propagator or the FPT distribution.

Finally, we note that the linear response in Eq. (\ref{gen_linear_response}) is the underlying MFPT squared, multiplied by a function of dimensionless parameters. Therefore, when studying the linear response of a process to a perturbation using experiments or numerical simulations, it is enough to scan the dimensionless parameters that characterize the system. This reduction of complexity is particularly useful when coming to study the response of a given process to various perturbations. Previously, doing so would have required dealing separately with each and every perturbation. This taxing procedure is no longer required as Eq. (\ref{gen_linear_response}) asserts that the response only depends on the perturbation via the normalized mean residual time to completion, $\overline{\tau}/\langle T \rangle$. Understanding how various perturbations influence this ratio is then the only remaining challenge.

\begin{acknowledgments}
T.D.K. acknowledges Shir Fridman for assistance in designing the figures. T.D.K. acknowledges insightful discussions with Ofek Lauber Bonomo. The authors acknowledge Maxence Arutkin, Haim Diamant, Barak Hirshberg, and Bara Levit for reading and commenting on early versions of this manuscript. This project has received funding from the European Research Council (ERC) under the European Union’s Horizon 2020 research and innovation program (grant agreement No. 947731).
\end{acknowledgments}

\appendix

\section{Derivation of Eq. (\ref{T_lambda MFPT TET})}\label{appendix: Eq. 5 Derivation}
To derive Eq. (\ref{T_lambda MFPT TET}), five terms are required. The probabilities $\Pr(T\leq P)$ and $\Pr(P<T)$, and the conditional expectations $\langle T|T\leq P\rangle$, $\langle P|P<T\rangle$, and $\langle \tau(\Vec{q})|P<T\rangle$. We start by computing the probabilities
\begin{equation}
\begin{split}
    \Pr(T\leq P)&=\int_0^\infty f_T(t)\Psi_P(t)\, dt=\\
    &=\int_0^\infty f_T(t)e^{-\lambda t}\, dt=\Tilde{T}(\lambda).\\
    \Pr(P<T)&=1-Pr(T\leq P)=1-\Tilde{T}(\lambda),
\end{split}
\end{equation}
where $f_X(t)$ and $\Psi_X(t)$ are the probability density function and the survival probability of the random variable $X$, respectively, and $\Tilde{T}(\lambda)$ is the Laplace transform of $f_T(t)$. The conditional expectations $\langle T|T\leq P\rangle$ and $\langle P|P<T\rangle$ are calculated as follows
\begin{equation}
    \begin{split}
        \langle T|T\leq P\rangle&=\frac{\int_0^\infty tf_T(t)\Psi_P(t)\, dt}{\Pr(T\leq P)}=\\
        &=\frac{-1}{\Tilde{T}(\lambda)}\frac{d}{d\lambda}\int_0^\infty f_T(t)e^{-\lambda t}\, dt=\\
        &=\frac{-1}{\Tilde{T}(\lambda)}\frac{d\Tilde{T}(\lambda)}{d\lambda}.\\
        \langle P|P<T\rangle&=\frac{\int_0^\infty tf_P(t)\Psi_T(t)\, dt}{\Pr(P<T)}=\\
        &=\frac{-\lambda}{1-\Tilde{T}(\lambda)}\frac{d}{d\lambda}\int_0^\infty \Psi_T(t)e^{-\lambda t}\, dt=\\
        &=\frac{1}{1-\Tilde{T}(\lambda)}\frac{d\Tilde{T}(\lambda)}{d\lambda}+\frac{1}{\lambda},
    \end{split}
\end{equation}
where we used the identity $\int_0^\infty \Psi_T(t)e^{-\lambda t}\, dt=\frac{1-\Tilde{T}(\lambda)}{\lambda}$.

Lastly, we need $\Pr(P<T)\langle \tau(\Vec{q})|P<T\rangle$. This can be done by averaging $\tau(\Vec{q})$ over all the realizations of $\tau(\Vec{q})$, and on the probability of being at state $\Vec{q}$ at the moment of the perturbation. We get
\begin{equation}\label{double integral}
\begin{split}
      \Pr(P<T)\langle \tau(\Vec{q})|P<T\rangle=\\=\lambda\int_Q\int_0^\infty e^{-\lambda t}\langle\tau(\Vec{q})\rangle G(\Vec{q},t)\, dt\, d\Vec{q}.
\end{split}
\end{equation}
Summing up all the terms we obtain Eq. (\ref{T_lambda MFPT TET}).

\section{Derivation of Eqs. (\ref{eq: lin approx}-\ref{gen_linear_response})}\label{appendix: dev of main res}
Eq. (\ref{T_lambda MFPT TET}) is a sum of two terms, $\frac{1-\Tilde{T}(\lambda)}{\lambda}$, and the right-hand side of Eq. (\ref{double integral}). We now expand both to first order in $\lambda$. For the first term, we use the moments expansion of the Laplace transform
\begin{equation}
    \frac{1-\Tilde{T}(\lambda)}{\lambda}=\langle T\rangle-\frac{\lambda}{2}\langle T^2\rangle+o(\lambda).
\end{equation}
For the right-hand side of Eq. (\ref{double integral}),
we assume that $\tau(\Vec{q})$ is completely determined given $\Vec{q}$, i.e., that it does not depend explicitly on the moment at which the perturbation is applied. Taking $\tau(\Vec{q})$ out of the inner integral, and Taylor expanding the exponential, we obtain
\begin{equation}
\begin{split}
    RHS&=\lambda\int_Q\langle\tau(\Vec{q})\rangle\int_0^\infty G(\Vec{q},t)\,dt\,d\Vec{q}+o(\lambda)=\\
&=\lambda A\int_Q\langle\tau(\Vec{q})\rangle\phi(\Vec{q})\,d\Vec{q}+o(\lambda)=\\
&=\lambda A\overline{\tau}+o(\lambda),
\end{split}
\end{equation}
where $\phi(\Vec{q})=A^{-1}\int_0^\infty G(\Vec{q},t)\,dt$ is the probability density created by integrating the propagator $G(\Vec{q},t)$ over all times, $A$ is the normalization factor, and $\overline{\tau}$ is the mean of $\langle\tau(\Vec{q})\rangle$ with respect to $\phi(\Vec{q})$. The value of $A$ is
\begin{equation}
\begin{split}
    A&=\int_Q\int_0^\infty G(\Vec{q},t)\, dt\, d\Vec{q} = \int_0^\infty\int_Q G(\Vec{q},t)\, d\Vec{q}\, dt=\\
    &=\int_0^\infty\Psi_T(t)\, dt = \langle T\rangle.
\end{split}
\end{equation}
Summing both results leads to Eqs. (\ref{eq: lin approx}-\ref{gen_linear_response}).

\section{Linear response in discrete time}\label{appendix: discrete time}
To derive the linear response for first-passage processes and perturbations occurring in discrete time, we return to Eq. (\ref{T_lambda definition}) in the main text. In this scenario, we consider a geometrically distributed perturbation time $P\sim \operatorname{Geometric}(p)$, and instead of $T_\lambda$ we denote the FPT of the perturbed process as $T_p$. Note, that in the case of discrete-time, the first-passage time is the number of steps taken to first-passage, and it is thus dimensionless. Using the total expectation theorem we get the following MFPT
\begin{equation}\label{MFPT DT}
    \langle T_p\rangle=\langle min(T,P)\rangle+\Pr(P<T)\langle\tau(\Vec{q})|P<T\rangle.
\end{equation}

\noindent We start by computing the first term
\begin{equation}
    \langle min(T,P)\rangle = \sum_{n=0}^\infty \Pr[min(T,P)>n],
\end{equation}
where $\Pr[min(T,P)>n]$ is the probability that both $T$ and $P$ are larger than $n$. Because $T$ and $P$ are independent
\begin{equation}\label{eq: minimal sur}
\begin{split}
    &\Pr[min(T,P)>n]=\\
    &=\left(\sum_{k=n+1}^\infty P_T(k)\right)\left(\sum_{m=n+1}^\infty P_P(m)\right)=\\
    &=\left(\sum_{k=n+1}^\infty P_T(k)\right)\left(\sum_{m=n+1}^\infty p(1-p)^{m}\right)=\\
    &=\sum_{k=n+1}^\infty P_T(k)(1-p)^{n+1}.
\end{split}
\end{equation}
Where $P_T(k)$ is the probability mass function of $T$, and $P_P(k)=p(1-p)^k$ is the probability mass function of $P$, where we used the fact that $P$ is geometrically distributed on $\{0,1,2,3,...\}$. To find $\langle min(T,P)\rangle$, we sum over Eq. (\ref{eq: minimal sur})
\begin{equation}
    \begin{split}
        \langle min(T,P)\rangle&=\sum_{n=0}^\infty \sum_{k=n+1}^\infty P_T(k)(1-p)^{n+1}=\\
        &=\sum_{k=1}^\infty P_T(k)\sum_{n=0}^{k-1}(1-p)^{n+1}=\\
        &=\frac{1-p}{p}\sum_{k=1}^\infty P_T(k)\left[1-(1-p)^{k}\right]=\\
        &=\frac{1-p}{p}\sum_{k=0}^\infty P_T(k)\left[1-(1-p)^{k}\right]=\\
        &=\frac{1-p}{p}\left[1-H_T(1-p)\right],
    \end{split}
\end{equation}
\noindent where $H_T(z)\equiv\sum_{n=0}^\infty P_T(n)z^n$ is the probability generating function of $T$. Similarly to Eq. (\ref{T_lambda MFPT TET}), we can see that the following result holds
\begin{equation}\label{DS better mean}
\begin{split}
    \langle T_p\rangle &= \frac{1-p}{p}\left(1-H_T(1-p)\right)+\\    &+p\int_Q\langle\tau(\Vec{q})\rangle\sum_{n=0}^{\infty}(1-p)^nG(\Vec{q},n)\,d\Vec{q}.
\end{split}
\end{equation}
In the limit of $p\to0$, we can find the linear response term, using the moments expansion of the probability generating function $H_T(z)$ around $z=1$. doing so, we obtain: $H_T(1-p)=1-p\langle T\rangle+\frac{p^2}{2}\langle T(T-1)\rangle+o(p^2)$. We will also define a probability density function $\phi(\Vec{q})$ similar to the one defined in Eq. (\ref{phi definition})
\begin{equation}
    \phi(\Vec{q})=\frac{1}{\langle T\rangle}\sum_{n=0}^\infty G(\Vec{q},n).
\end{equation}
Using this definition of $\phi(\Vec{q})$ we can construct the discrete-time analog of $\overline{\tau}$ according to Eq. (\ref{tau bar definition}). Doing so gives the following expression for the linear response
\begin{equation}\label{eq: discrete-time gen linear response}
    \chi = \frac{\langle T\rangle^2}{2}\left(\frac{2\overline{\tau}}{\langle T\rangle}-CV_T^2-1-\frac{1}{\langle T\rangle}\right).
\end{equation}
In the continuous limit, i.e., $\langle T\rangle\gg 1$, it is easy to see that Eq. (\ref{eq: discrete-time gen linear response}) converges to Eq. (\ref{gen_linear_response}), as expected. 

For stochastic resetting, $\overline{\tau}=\langle T\rangle$. Therefore, $\chi=\langle T\rangle^2\left(1-CV_T^2-\langle T\rangle^{-1}\right)/2$, which agrees with the known result for stochastic resetting in discrete time \cite{StocResetDT}.

\section{MFPT of informed resetting when $\sigma/L=0$}\label{appendix: full information}

The case of full information, i.e., $\sigma/L=0$ was solved in \cite{Asymmetric_SR} for overdamped motion obeying the Fokker-Planck equation. Using Eq. 12 in \cite{Asymmetric_SR}, one can see that to find the MFPT, the following ordinary differential equation should be solved
\begin{equation}
\begin{split}
    &\frac{d^2\langle T_\lambda(x)\rangle}{dx^2}-\frac{v}{D}\frac{d\langle T_\lambda(x)\rangle}{dx}+\\
    &+\frac{\lambda}{D}\Theta(x-L)\left[\langle T_\lambda(L)\rangle-\langle T_\lambda(x)\rangle\right]=-\frac{1}{D},
\end{split}
\end{equation}
where $L$ is the distance of the resetting point from the absorbing boundary, $v$ is the drift velocity, $D$ is the diffusion coefficient, $\lambda$ is the informed resetting rate, and $\Theta(y)$ is the Heaviside function. The resulting function $\langle T_\lambda(x)\rangle$ is the MFPT from an initial condition distanced $x$ from the absorbing boundary. Therefore the boundary conditions are $\langle T_\lambda(0)\rangle=0$ and $0\leq\lim_{x\to\infty}\langle T_\lambda(x)\rangle<\infty$. Moreover, both $\langle T_\lambda(x)\rangle$ and its first derivative must be continuous at $x=L$. 

The MFPT for our problem is $\langle T_\lambda(L)\rangle$. To get it, we solve for particles initialized from $0<x<L$, and using the boundary condition at $x=0$ gives the following solution
\begin{equation}
    \langle T_\lambda(x)\rangle_{x<L}=\frac{x}{v}+A\left(1-e^{\frac{v}{D}x}\right),
\end{equation}
where $A$ is some constant that will be found using the continuity of the first derivative at $x=L$. For $x>L$ the solution is
\begin{equation}\label{Eq: D3}
    \langle T_\lambda(x)\rangle_{x>L}=\langle T_\lambda(L)\rangle+\frac{1}{\lambda}\left(1-e^{\frac{v}{2D}\left[1-\alpha(\lambda)\right](x-L)}\right),
\end{equation}
where $\alpha(\lambda)\equiv\sqrt{1+\frac{4D\lambda}{v^2}}$. Using the continuity of the first derivative at $x=L$, we find $A=\left(\frac{D}{v^2}+\frac{1-\alpha(\lambda)}{2\lambda}\right)e^{-2Pe}$. Having found $A$, we can evaluate $\langle T_\lambda(L)\rangle$, and get the desired MFPT
\begin{equation}
    \langle T_\lambda\rangle=\frac{L}{v}+\left(\frac{D}{v^2}+\frac{1-\alpha(\lambda)}{2\lambda}\right)\left(e^{-2Pe}-1\right).
\end{equation}
Taking the derivative with respect to $\lambda$ of the equation above at $\lambda=0$ agrees with taking the limit of $\sigma/L\to0$ in Eq. (\ref{linear response informed}).

\section{MFPT for stochastic field activation}\label{appendix: field activation}

As discussed in the main text, Eq. (\ref{T_lambda MFPT TET}) holds for an arbitrary perturbation rate. Moreover, in the case of stochastic field activation, the field is activated once and for all, eliminating the $\lambda$ dependency of $\langle\tau(\Vec{q})\rangle$. Therefore, we can simplify Eq. (\ref{T_lambda MFPT TET}) to read
\begin{equation}\label{MFPT for one time pert}
    \langle T_\lambda\rangle=\frac{1-\Tilde{T}(\lambda)}{\lambda}+\lambda\int_Q \langle \tau(\Vec{q})\rangle \Tilde{G}(\Vec{q},\lambda)\, d\Vec{q},
\end{equation}

\noindent where $\Tilde{G}(\Vec{q},\lambda)$ is the Laplace transform of $G(\Vec{q},t)$ evaluated at $\lambda$. 

In the case of a drift-diffusion process, where the perturbation is activating a field that alters the potential to a logarithmic one, all terms in Eq. (\ref{MFPT for one time pert}) are known. $\Tilde{T}(\lambda)$ is the Laplace transform of the FPT distribution of drift-diffusion which is $\Tilde{T}(\lambda)=\exp{\left[Pe\left(1-\alpha(\lambda)\right)\right]}$ \cite{redner_book}, $\langle\tau(\Vec{q})\rangle$ is given by Eq. (\ref{log MFPT}) and $\Tilde{G}(\Vec{q},t)$  is the Laplace transform of the drift-diffusion propagator \cite{redner_book}
\begin{equation}
    \Tilde{G}(x,\lambda)=\frac{e^{Pe\left(\frac{x}{L}-\alpha(\lambda)\frac{|x|}{L}\right)}-e^{Pe\left[\frac{x}{L}+\alpha(\lambda)\left(\frac{x}{L}-2\right)\right]}}{v\alpha(\lambda)},
\end{equation}
where $\alpha(\lambda)$ was defined in Eq. (\ref{Eq: D3}). Inserting all those results into Eq. (\ref{MFPT for one time pert}), gives the following MFPT
\begin{equation}\label{eq: log gen MFPT}
\begin{split}
    \langle T_\lambda\rangle&=\frac{L^2}{2D(\beta U_0-1)}-\frac{2Pe}{\lambda(\beta U_0-1)}+\\
    &+\left[\frac{v^2}{\lambda D}+\beta U_0\right]\frac{1-e^{Pe[1-\alpha(\lambda)]}}{\lambda(\beta U_0-1)}.
\end{split}
\end{equation}
Substituting $\lambda=0$ into the above equation gives $\langle T_\lambda\rangle=L/v$, and its derivative at this point is given by Eq. (\ref{linear response change of potential}), as expected. For the case where $\beta U_0<1$ the MFPT on the logarithmic potential diverges, and therefore $\langle T_\lambda\rangle$ in that case is
\begin{equation}
    \langle T_\lambda\rangle = \begin{cases}
        \frac{L}{v}&\text{if $\lambda=0$,}\\
        \infty &\text{if $\lambda>0$.}
    \end{cases}
\end{equation}

\subsection{Scaling of  $\lambda^\star$ near the transition between the finite and infinite phases}

In the general case, finding $\lambda^\star$ requires finding the $\lambda$ that minimizes Eq. (\ref{eq: log gen MFPT}). This task boils down to solving a transcendental equation, which can be done only numerically. However, in the transition between the finite to infinite $\lambda^\star$ phases, Eq. (\ref{eq: log gen MFPT}) can be simplified, because we know that the minimum occurs at $\lambda\gg v^2/D$.

Therefore, in this regime $e^{Pe[1-\alpha(\lambda)]}\ll 1$, and the derivative of Eq. (\ref{eq: log gen MFPT}) is approximately given by
\begin{equation}
    \frac{d\langle T_\lambda\rangle}{d\lambda}\simeq \frac{2Pe-\beta U_0}{\lambda^2(\beta U_0-1)} -\frac{2v^2}{\lambda^3D(\beta U_0-1)}.
\end{equation}
Equating this derivative to zero, we find that in the finite $\lambda^\star$ phase, close to the transition to the infinite $\lambda^\star$ phase, we have 
\begin{equation}
    \lambda^\star\simeq \frac{2v^2}{D(2Pe-\beta U_0)}\sim(2Pe-\beta U_0)^{-1}.
\end{equation}

\section{Details of molecular degradation simulations in Fig. 7}\label{appendix: simulation details}
We simulated two Markov models for the degradation process: a linear model and a parallel model (Fig. \ref{fig: little law}c). In both cases, synthesis was taken to be a Poisson process with rate $11/21$. The transition rates for the linear model were set to $k_1=k_2=22/21$, and for the parallel model they were set to: $k_1=1,\,k_2=10,\,k_3=0.1,\,k_4=0.1$. The lifetime of a molecule is the minimal value between the native degradation time, and an exponentially distributed time with rate $k_\text{on}[A]$, accounting for degradation by the annihilator. Each symbol in Fig. (\ref{fig: little law}d) is the average, over $2\cdot10^8$ realizations, of the concentration at time $t=100$, which represents the steady-state concentration.

For Fig. (\ref{fig: little law}e) we used $2\cdot10^7$ realizations, at thirty annihilator concentrations linearly spaced between $10^{-6} \,k_{\text{on}}^{-1}$ and $10\, k_{\text{on}}^{-1}$. We used a Tikhonov regularized inverse Laplace transform \cite{Istratov1999}; the regularization was done using the second derivative of the distribution in $t$-space. The regularization factor was selected using the generalized cross-validation method \cite{Golub1979}. The inversion was done to $10^4$ points in $t$-space, linearly spaced between $0$ and $20$.

\bibliography{bibliography}

\begin{thebibliography}{72}%
\makeatletter
\providecommand \@ifxundefined [1]{%
 \@ifx{#1\undefined}
}%
\providecommand \@ifnum [1]{%
 \ifnum #1\expandafter \@firstoftwo
 \else \expandafter \@secondoftwo
 \fi
}%
\providecommand \@ifx [1]{%
 \ifx #1\expandafter \@firstoftwo
 \else \expandafter \@secondoftwo
 \fi
}%
\providecommand \natexlab [1]{#1}%
\providecommand \enquote  [1]{``#1''}%
\providecommand \bibnamefont  [1]{#1}%
\providecommand \bibfnamefont [1]{#1}%
\providecommand \citenamefont [1]{#1}%
\providecommand \href@noop [0]{\@secondoftwo}%
\providecommand \href [0]{\begingroup \@sanitize@url \@href}%
\providecommand \@href[1]{\@@startlink{#1}\@@href}%
\providecommand \@@href[1]{\endgroup#1\@@endlink}%
\providecommand \@sanitize@url [0]{\catcode `\\12\catcode `\$12\catcode
  `\&12\catcode `\#12\catcode `\^12\catcode `\_12\catcode `\%12\relax}%
\providecommand \@@startlink[1]{}%
\providecommand \@@endlink[0]{}%
\providecommand \url  [0]{\begingroup\@sanitize@url \@url }%
\providecommand \@url [1]{\endgroup\@href {#1}{\urlprefix }}%
\providecommand \urlprefix  [0]{URL }%
\providecommand \Eprint [0]{\href }%
\providecommand \doibase [0]{https://doi.org/}%
\providecommand \selectlanguage [0]{\@gobble}%
\providecommand \bibinfo  [0]{\@secondoftwo}%
\providecommand \bibfield  [0]{\@secondoftwo}%
\providecommand \translation [1]{[#1]}%
\providecommand \BibitemOpen [0]{}%
\providecommand \bibitemStop [0]{}%
\providecommand \bibitemNoStop [0]{.\EOS\space}%
\providecommand \EOS [0]{\spacefactor3000\relax}%
\providecommand \BibitemShut  [1]{\csname bibitem#1\endcsname}%
\let\auto@bib@innerbib\@empty
%</preamble>
\bibitem [{\citenamefont {Marconi}\ \emph {et~al.}(2008)\citenamefont
  {Marconi}, \citenamefont {Puglisi}, \citenamefont {Rondoni},\ and\
  \citenamefont {Vulpiani}}]{FDR_review}%
  \BibitemOpen
  \bibfield  {author} {\bibinfo {author} {\bibfnamefont {U.~M.~B.}\
  \bibnamefont {Marconi}}, \bibinfo {author} {\bibfnamefont {A.}~\bibnamefont
  {Puglisi}}, \bibinfo {author} {\bibfnamefont {L.}~\bibnamefont {Rondoni}},\
  and\ \bibinfo {author} {\bibfnamefont {A.}~\bibnamefont {Vulpiani}},\
  }\bibfield  {title} {\bibinfo {title} {Fluctuation–dissipation: Response
  theory in statistical physics},\ }\href@noop {} {\bibfield  {journal}
  {\bibinfo  {journal} {Physics Reports}\ }\textbf {\bibinfo {volume} {461}},\
  \bibinfo {pages} {111} (\bibinfo {year} {2008})}\BibitemShut {NoStop}%
\bibitem [{\citenamefont {Einstein}(1905)}]{Einstein_BM}%
  \BibitemOpen
  \bibfield  {author} {\bibinfo {author} {\bibfnamefont {A.}~\bibnamefont
  {Einstein}},\ }\bibfield  {title} {\bibinfo {title} {Über die von der
  molekularkinetischen theorie der wärme geforderte bewegung von in ruhenden
  flüssigkeiten suspendierten teilchen},\ }\href@noop {} {\bibfield  {journal}
  {\bibinfo  {journal} {Annalen der Physik}\ }\textbf {\bibinfo {volume}
  {322}},\ \bibinfo {pages} {549} (\bibinfo {year} {1905})}\BibitemShut
  {NoStop}%
\bibitem [{\citenamefont {Johnson}(1928)}]{Johnson}%
  \BibitemOpen
  \bibfield  {author} {\bibinfo {author} {\bibfnamefont {J.~B.}\ \bibnamefont
  {Johnson}},\ }\bibfield  {title} {\bibinfo {title} {Thermal agitation of
  electricity in conductors},\ }\href@noop {} {\bibfield  {journal} {\bibinfo
  {journal} {Phys. Rev.}\ }\textbf {\bibinfo {volume} {32}},\ \bibinfo {pages}
  {97} (\bibinfo {year} {1928})}\BibitemShut {NoStop}%
\bibitem [{\citenamefont {Nyquist}(1928)}]{Nyquist}%
  \BibitemOpen
  \bibfield  {author} {\bibinfo {author} {\bibfnamefont {H.}~\bibnamefont
  {Nyquist}},\ }\bibfield  {title} {\bibinfo {title} {Thermal agitation of
  electric charge in conductors},\ }\href@noop {} {\bibfield  {journal}
  {\bibinfo  {journal} {Phys. Rev.}\ }\textbf {\bibinfo {volume} {32}},\
  \bibinfo {pages} {110} (\bibinfo {year} {1928})}\BibitemShut {NoStop}%
\bibitem [{\citenamefont {Redner}(2001)}]{redner_book}%
  \BibitemOpen
  \bibfield  {author} {\bibinfo {author} {\bibfnamefont {S.}~\bibnamefont
  {Redner}},\ }\href@noop {} {\emph {\bibinfo {title} {A Guide to First-Passage
  Processes}}}\ (\bibinfo  {publisher} {Cambridge University Press},\ \bibinfo
  {address} {United Kingdom},\ \bibinfo {year} {2001})\BibitemShut {NoStop}%
\bibitem [{\citenamefont {Alan J.~Bray}\ and\ \citenamefont
  {Schehr}(2013)}]{FP_review_Bray_Majumdar_Scher}%
  \BibitemOpen
  \bibfield  {author} {\bibinfo {author} {\bibfnamefont {S.~N.~M.}\
  \bibnamefont {Alan J.~Bray}}\ and\ \bibinfo {author} {\bibfnamefont
  {G.}~\bibnamefont {Schehr}},\ }\bibfield  {title} {\bibinfo {title}
  {Persistence and first-passage properties in nonequilibrium systems},\
  }\href@noop {} {\bibfield  {journal} {\bibinfo  {journal} {Advances in
  Physics}\ }\textbf {\bibinfo {volume} {62}},\ \bibinfo {pages} {225}
  (\bibinfo {year} {2013})}\BibitemShut {NoStop}%
\bibitem [{\citenamefont {Metzler}\ \emph {et~al.}(2014)\citenamefont
  {Metzler}, \citenamefont {Oshanin},\ and\ \citenamefont
  {Redner}}]{FP_and_applications}%
  \BibitemOpen
  \bibfield  {author} {\bibinfo {author} {\bibfnamefont {R.}~\bibnamefont
  {Metzler}}, \bibinfo {author} {\bibfnamefont {G.}~\bibnamefont {Oshanin}},\
  and\ \bibinfo {author} {\bibfnamefont {S.}~\bibnamefont {Redner}},\
  }\href@noop {} {\emph {\bibinfo {title} {First-Passage Phenomena and Their
  Applications}}}\ (\bibinfo  {publisher} {World Scientific},\ \bibinfo
  {address} {Singapore},\ \bibinfo {year} {2014})\BibitemShut {NoStop}%
\bibitem [{\citenamefont {Grebenkov}\ \emph {et~al.}(2024)\citenamefont
  {Grebenkov}, \citenamefont {Metzler},\ and\ \citenamefont
  {Oshanin}}]{Target_Search_Problems}%
  \BibitemOpen
  \bibfield  {author} {\bibinfo {author} {\bibfnamefont {D.}~\bibnamefont
  {Grebenkov}}, \bibinfo {author} {\bibfnamefont {R.}~\bibnamefont {Metzler}},\
  and\ \bibinfo {author} {\bibfnamefont {G.~G.}\ \bibnamefont {Oshanin}},\
  }\href@noop {} {\emph {\bibinfo {title} {Target Search Problems}}},\ \bibinfo
  {edition} {first edition.}\ ed.\ (\bibinfo  {publisher} {Springer},\ \bibinfo
  {address} {Cham, Switzerland},\ \bibinfo {year} {2024})\BibitemShut {NoStop}%
\bibitem [{\citenamefont {Giuggioli}(2020)}]{LRW_in_arb_dimensions}%
  \BibitemOpen
  \bibfield  {author} {\bibinfo {author} {\bibfnamefont {L.}~\bibnamefont
  {Giuggioli}},\ }\bibfield  {title} {\bibinfo {title} {Exact spatiotemporal
  dynamics of confined lattice random walks in arbitrary dimensions: A century
  after smoluchowski and p\'olya},\ }\href@noop {} {\bibfield  {journal}
  {\bibinfo  {journal} {Phys. Rev. X}\ }\textbf {\bibinfo {volume} {10}},\
  \bibinfo {pages} {021045} (\bibinfo {year} {2020})}\BibitemShut {NoStop}%
\bibitem [{\citenamefont {Engelberg}\ \emph {et~al.}(2018)\citenamefont
  {Engelberg}, \citenamefont {Ashkenazy},\ and\ \citenamefont
  {Assaf}}]{Stochastic_BD}%
  \BibitemOpen
  \bibfield  {author} {\bibinfo {author} {\bibfnamefont {E.~Z.}\ \bibnamefont
  {Engelberg}}, \bibinfo {author} {\bibfnamefont {Y.}~\bibnamefont
  {Ashkenazy}},\ and\ \bibinfo {author} {\bibfnamefont {M.}~\bibnamefont
  {Assaf}},\ }\bibfield  {title} {\bibinfo {title} {Stochastic model of
  breakdown nucleation under intense electric fields},\ }\href@noop {}
  {\bibfield  {journal} {\bibinfo  {journal} {Phys. Rev. Lett.}\ }\textbf
  {\bibinfo {volume} {120}},\ \bibinfo {pages} {124801} (\bibinfo {year}
  {2018})}\BibitemShut {NoStop}%
\bibitem [{\citenamefont {Gu{\'e}rin}\ \emph {et~al.}(2016)\citenamefont
  {Gu{\'e}rin}, \citenamefont {Levernier}, \citenamefont {B{\'e}nichou},\ and\
  \citenamefont {Voituriez}}]{Non_Markov_MFPT}%
  \BibitemOpen
  \bibfield  {author} {\bibinfo {author} {\bibfnamefont {T.}~\bibnamefont
  {Gu{\'e}rin}}, \bibinfo {author} {\bibfnamefont {N.}~\bibnamefont
  {Levernier}}, \bibinfo {author} {\bibfnamefont {O.}~\bibnamefont
  {B{\'e}nichou}},\ and\ \bibinfo {author} {\bibfnamefont {R.}~\bibnamefont
  {Voituriez}},\ }\bibfield  {title} {\bibinfo {title} {Mean first-passage
  times of non-markovian random walkers in confinement},\ }\href@noop {}
  {\bibfield  {journal} {\bibinfo  {journal} {Nature}\ }\textbf {\bibinfo
  {volume} {534}},\ \bibinfo {pages} {356} (\bibinfo {year}
  {2016})}\BibitemShut {NoStop}%
\bibitem [{\citenamefont {Bernardi}\ and\ \citenamefont
  {Lindner}(2022)}]{Run_with_brown}%
  \BibitemOpen
  \bibfield  {author} {\bibinfo {author} {\bibfnamefont {D.}~\bibnamefont
  {Bernardi}}\ and\ \bibinfo {author} {\bibfnamefont {B.}~\bibnamefont
  {Lindner}},\ }\bibfield  {title} {\bibinfo {title} {Run with the brownian
  hare, hunt with the deterministic hounds},\ }\href@noop {} {\bibfield
  {journal} {\bibinfo  {journal} {Phys. Rev. Lett.}\ }\textbf {\bibinfo
  {volume} {128}},\ \bibinfo {pages} {040601} (\bibinfo {year}
  {2022})}\BibitemShut {NoStop}%
\bibitem [{\citenamefont {Condamin}\ \emph {et~al.}(2007)\citenamefont
  {Condamin}, \citenamefont {B{\'e}nichou}, \citenamefont {Tejedor},
  \citenamefont {Voituriez},\ and\ \citenamefont {Klafter}}]{condamin2007}%
  \BibitemOpen
  \bibfield  {author} {\bibinfo {author} {\bibfnamefont {S.}~\bibnamefont
  {Condamin}}, \bibinfo {author} {\bibfnamefont {O.}~\bibnamefont
  {B{\'e}nichou}}, \bibinfo {author} {\bibfnamefont {V.}~\bibnamefont
  {Tejedor}}, \bibinfo {author} {\bibfnamefont {R.}~\bibnamefont {Voituriez}},\
  and\ \bibinfo {author} {\bibfnamefont {J.}~\bibnamefont {Klafter}},\
  }\bibfield  {title} {\bibinfo {title} {First-passage times in complex
  scale-invariant media},\ }\href@noop {} {\bibfield  {journal} {\bibinfo
  {journal} {Nature}\ }\textbf {\bibinfo {volume} {450}},\ \bibinfo {pages}
  {77} (\bibinfo {year} {2007})}\BibitemShut {NoStop}%
\bibitem [{\citenamefont {Thorneywork}\ \emph {et~al.}(2020)\citenamefont
  {Thorneywork}, \citenamefont {Gladrow}, \citenamefont {Qing}, \citenamefont
  {Rico-Pasto}, \citenamefont {Ritort}, \citenamefont {Bayley}, \citenamefont
  {Kolomeisky},\ and\ \citenamefont {Keyser}}]{Energy_landscape_FPT}%
  \BibitemOpen
  \bibfield  {author} {\bibinfo {author} {\bibfnamefont {A.~L.}\ \bibnamefont
  {Thorneywork}}, \bibinfo {author} {\bibfnamefont {J.}~\bibnamefont
  {Gladrow}}, \bibinfo {author} {\bibfnamefont {Y.}~\bibnamefont {Qing}},
  \bibinfo {author} {\bibfnamefont {M.}~\bibnamefont {Rico-Pasto}}, \bibinfo
  {author} {\bibfnamefont {F.}~\bibnamefont {Ritort}}, \bibinfo {author}
  {\bibfnamefont {H.}~\bibnamefont {Bayley}}, \bibinfo {author} {\bibfnamefont
  {A.~B.}\ \bibnamefont {Kolomeisky}},\ and\ \bibinfo {author} {\bibfnamefont
  {U.~F.}\ \bibnamefont {Keyser}},\ }\bibfield  {title} {\bibinfo {title}
  {Direct detection of molecular intermediates from first-passage times},\
  }\href@noop {} {\bibfield  {journal} {\bibinfo  {journal} {Science Advances}\
  }\textbf {\bibinfo {volume} {6}},\ \bibinfo {pages} {eaaz4642} (\bibinfo
  {year} {2020})}\BibitemShut {NoStop}%
\bibitem [{\citenamefont {Lanoiselée}\ \emph {et~al.}(2018)\citenamefont
  {Lanoiselée}, \citenamefont {Moutal},\ and\ \citenamefont
  {Grebenkov}}]{Diffusing_Diffusivities_limted_reactions}%
  \BibitemOpen
  \bibfield  {author} {\bibinfo {author} {\bibfnamefont {Y.}~\bibnamefont
  {Lanoiselée}}, \bibinfo {author} {\bibfnamefont {N.}~\bibnamefont
  {Moutal}},\ and\ \bibinfo {author} {\bibfnamefont {D.~S.}\ \bibnamefont
  {Grebenkov}},\ }\bibfield  {title} {\bibinfo {title} {Diffusion-limited
  reactions in dynamic heterogeneous media},\ }\href@noop {} {\bibfield
  {journal} {\bibinfo  {journal} {Nature Communications}\ }\textbf {\bibinfo
  {volume} {9}},\ \bibinfo {pages} {4398} (\bibinfo {year} {2018})}\BibitemShut
  {NoStop}%
\bibitem [{\citenamefont {Sokolov}(2003)}]{Sokolov2003Cyclization}%
  \BibitemOpen
  \bibfield  {author} {\bibinfo {author} {\bibfnamefont {I.~M.}\ \bibnamefont
  {Sokolov}},\ }\bibfield  {title} {\bibinfo {title} {Cyclization of a polymer:
  First-passage problem for a non-markovian process},\ }\href@noop {}
  {\bibfield  {journal} {\bibinfo  {journal} {Phys. Rev. Lett.}\ }\textbf
  {\bibinfo {volume} {90}},\ \bibinfo {pages} {080601} (\bibinfo {year}
  {2003})}\BibitemShut {NoStop}%
\bibitem [{\citenamefont {Borberg}\ \emph {et~al.}(2019)\citenamefont
  {Borberg}, \citenamefont {Zverzhinetsky}, \citenamefont {Krivitsky},
  \citenamefont {Kosloff}, \citenamefont {Heifler}, \citenamefont {Degabli},
  \citenamefont {Soroka}, \citenamefont {Fainaro}, \citenamefont {Burstein},
  \citenamefont {Reuveni}, \citenamefont {Diamant}, \citenamefont {Krivitsky},\
  and\ \citenamefont {Patolsky}}]{Collect_realse}%
  \BibitemOpen
  \bibfield  {author} {\bibinfo {author} {\bibfnamefont {E.}~\bibnamefont
  {Borberg}}, \bibinfo {author} {\bibfnamefont {M.}~\bibnamefont
  {Zverzhinetsky}}, \bibinfo {author} {\bibfnamefont {A.}~\bibnamefont
  {Krivitsky}}, \bibinfo {author} {\bibfnamefont {A.}~\bibnamefont {Kosloff}},
  \bibinfo {author} {\bibfnamefont {O.}~\bibnamefont {Heifler}}, \bibinfo
  {author} {\bibfnamefont {G.}~\bibnamefont {Degabli}}, \bibinfo {author}
  {\bibfnamefont {H.~P.}\ \bibnamefont {Soroka}}, \bibinfo {author}
  {\bibfnamefont {R.~S.}\ \bibnamefont {Fainaro}}, \bibinfo {author}
  {\bibfnamefont {L.}~\bibnamefont {Burstein}}, \bibinfo {author}
  {\bibfnamefont {S.}~\bibnamefont {Reuveni}}, \bibinfo {author} {\bibfnamefont
  {H.}~\bibnamefont {Diamant}}, \bibinfo {author} {\bibfnamefont
  {V.}~\bibnamefont {Krivitsky}},\ and\ \bibinfo {author} {\bibfnamefont
  {F.}~\bibnamefont {Patolsky}},\ }\bibfield  {title} {\bibinfo {title}
  {Light-controlled selective collection-and-release of biomolecules by an
  on-chip nanostructured device},\ }\href@noop {} {\bibfield  {journal}
  {\bibinfo  {journal} {Nano Letters}\ }\textbf {\bibinfo {volume} {19}},\
  \bibinfo {pages} {5868} (\bibinfo {year} {2019})}\BibitemShut {NoStop}%
\bibitem [{\citenamefont {Scher}\ \emph {et~al.}(2023)\citenamefont {Scher},
  \citenamefont {Reuveni},\ and\ \citenamefont {Grebenkov}}]{Sticky_particle}%
  \BibitemOpen
  \bibfield  {author} {\bibinfo {author} {\bibfnamefont {Y.}~\bibnamefont
  {Scher}}, \bibinfo {author} {\bibfnamefont {S.}~\bibnamefont {Reuveni}},\
  and\ \bibinfo {author} {\bibfnamefont {D.~S.}\ \bibnamefont {Grebenkov}},\
  }\bibfield  {title} {\bibinfo {title} {Escape of a sticky particle},\
  }\href@noop {} {\bibfield  {journal} {\bibinfo  {journal} {Phys. Rev. Res.}\
  }\textbf {\bibinfo {volume} {5}},\ \bibinfo {pages} {043196} (\bibinfo {year}
  {2023})}\BibitemShut {NoStop}%
\bibitem [{\citenamefont {Sokolov}\ \emph {et~al.}(2005)\citenamefont
  {Sokolov}, \citenamefont {Metzler}, \citenamefont {Pant},\ and\ \citenamefont
  {Williams}}]{Search_on_DNA}%
  \BibitemOpen
  \bibfield  {author} {\bibinfo {author} {\bibfnamefont {I.~M.}\ \bibnamefont
  {Sokolov}}, \bibinfo {author} {\bibfnamefont {R.}~\bibnamefont {Metzler}},
  \bibinfo {author} {\bibfnamefont {K.}~\bibnamefont {Pant}},\ and\ \bibinfo
  {author} {\bibfnamefont {M.~C.}\ \bibnamefont {Williams}},\ }\bibfield
  {title} {\bibinfo {title} {Target search of n sliding proteins on a dna},\
  }\href@noop {} {\bibfield  {journal} {\bibinfo  {journal} {Biophysical
  Journal}\ }\textbf {\bibinfo {volume} {89}},\ \bibinfo {pages} {895}
  (\bibinfo {year} {2005})}\BibitemShut {NoStop}%
\bibitem [{\citenamefont {Gallos}\ \emph {et~al.}(2007)\citenamefont {Gallos},
  \citenamefont {Song}, \citenamefont {Havlin},\ and\ \citenamefont
  {Makse}}]{Transport_on_fractal_networks}%
  \BibitemOpen
  \bibfield  {author} {\bibinfo {author} {\bibfnamefont {L.~K.}\ \bibnamefont
  {Gallos}}, \bibinfo {author} {\bibfnamefont {C.}~\bibnamefont {Song}},
  \bibinfo {author} {\bibfnamefont {S.}~\bibnamefont {Havlin}},\ and\ \bibinfo
  {author} {\bibfnamefont {H.~A.}\ \bibnamefont {Makse}},\ }\bibfield  {title}
  {\bibinfo {title} {Scaling theory of transport in complex biological
  networks},\ }\href@noop {} {\bibfield  {journal} {\bibinfo  {journal}
  {Proceedings of the National Academy of Sciences}\ }\textbf {\bibinfo
  {volume} {104}},\ \bibinfo {pages} {7746} (\bibinfo {year}
  {2007})}\BibitemShut {NoStop}%
\bibitem [{\citenamefont {Eliazar}\ \emph {et~al.}(2007)\citenamefont
  {Eliazar}, \citenamefont {Koren},\ and\ \citenamefont
  {Klafter}}]{Circular_DNA_Search}%
  \BibitemOpen
  \bibfield  {author} {\bibinfo {author} {\bibfnamefont {I.}~\bibnamefont
  {Eliazar}}, \bibinfo {author} {\bibfnamefont {T.}~\bibnamefont {Koren}},\
  and\ \bibinfo {author} {\bibfnamefont {J.}~\bibnamefont {Klafter}},\
  }\bibfield  {title} {\bibinfo {title} {Searching circular dna strands},\
  }\href@noop {} {\bibfield  {journal} {\bibinfo  {journal} {Journal of
  Physics: Condensed Matter}\ }\textbf {\bibinfo {volume} {19}},\ \bibinfo
  {pages} {065140} (\bibinfo {year} {2007})}\BibitemShut {NoStop}%
\bibitem [{\citenamefont {Meyer}\ and\ \citenamefont
  {Rieger}(2021)}]{n_steps_memory}%
  \BibitemOpen
  \bibfield  {author} {\bibinfo {author} {\bibfnamefont {H.}~\bibnamefont
  {Meyer}}\ and\ \bibinfo {author} {\bibfnamefont {H.}~\bibnamefont {Rieger}},\
  }\bibfield  {title} {\bibinfo {title} {Optimal non-markovian search
  strategies with $n$-step memory},\ }\href@noop {} {\bibfield  {journal}
  {\bibinfo  {journal} {Phys. Rev. Lett.}\ }\textbf {\bibinfo {volume} {127}},\
  \bibinfo {pages} {070601} (\bibinfo {year} {2021})}\BibitemShut {NoStop}%
\bibitem [{\citenamefont {Schwarz}\ \emph {et~al.}(2016)\citenamefont
  {Schwarz}, \citenamefont {Schr\"oder}, \citenamefont {Qu}, \citenamefont
  {Hoth},\ and\ \citenamefont {Rieger}}]{Spat_inhom_search}%
  \BibitemOpen
  \bibfield  {author} {\bibinfo {author} {\bibfnamefont {K.}~\bibnamefont
  {Schwarz}}, \bibinfo {author} {\bibfnamefont {Y.}~\bibnamefont {Schr\"oder}},
  \bibinfo {author} {\bibfnamefont {B.}~\bibnamefont {Qu}}, \bibinfo {author}
  {\bibfnamefont {M.}~\bibnamefont {Hoth}},\ and\ \bibinfo {author}
  {\bibfnamefont {H.}~\bibnamefont {Rieger}},\ }\bibfield  {title} {\bibinfo
  {title} {Optimality of spatially inhomogeneous search strategies},\
  }\href@noop {} {\bibfield  {journal} {\bibinfo  {journal} {Phys. Rev. Lett.}\
  }\textbf {\bibinfo {volume} {117}},\ \bibinfo {pages} {068101} (\bibinfo
  {year} {2016})}\BibitemShut {NoStop}%
\bibitem [{\citenamefont {Iyer-Biswas}\ and\ \citenamefont
  {Zilman}(2016)}]{FP_in_cell_biology}%
  \BibitemOpen
  \bibfield  {author} {\bibinfo {author} {\bibfnamefont {S.}~\bibnamefont
  {Iyer-Biswas}}\ and\ \bibinfo {author} {\bibfnamefont {A.}~\bibnamefont
  {Zilman}},\ }\bibfield  {title} {\bibinfo {title} {First-passage processes in
  cellular biology},\ }\href@noop {} {\bibfield  {journal} {\bibinfo  {journal}
  {Advances in Chemical Physics}\ } (\bibinfo {year} {2016})}\BibitemShut
  {NoStop}%
\bibitem [{\citenamefont {Kochugaeva}\ \emph {et~al.}()\citenamefont
  {Kochugaeva}, \citenamefont {Shvets},\ and\ \citenamefont
  {Kolomeisky}}]{Chemical_kinetics_beyond}%
  \BibitemOpen
  \bibfield  {author} {\bibinfo {author} {\bibfnamefont {M.~P.}\ \bibnamefont
  {Kochugaeva}}, \bibinfo {author} {\bibfnamefont {A.~A.}\ \bibnamefont
  {Shvets}},\ and\ \bibinfo {author} {\bibfnamefont {A.~B.}\ \bibnamefont
  {Kolomeisky}},\ }\bibinfo {title} {Kinetics of protein–dna interactions:
  First-passage analysis},\ in\ \href@noop {} {\emph {\bibinfo {booktitle}
  {Chemical Kinetics}}},\ Chap.~\bibinfo {chapter} {19}, pp.\ \bibinfo {pages}
  {473--492}\BibitemShut {NoStop}%
\bibitem [{\citenamefont {Chicheportiche}\ and\ \citenamefont
  {Bouchaud}(2014)}]{JPBouchaud_book_chapter}%
  \BibitemOpen
  \bibfield  {author} {\bibinfo {author} {\bibfnamefont {R.}~\bibnamefont
  {Chicheportiche}}\ and\ \bibinfo {author} {\bibfnamefont {J.-P.}\
  \bibnamefont {Bouchaud}},\ }\bibinfo {title} {Some applications of
  first-passage ideas to finance},\ in\ \href@noop {} {\emph {\bibinfo
  {booktitle} {First-Passage Phenomena and Their Applications}}}\ (\bibinfo
  {publisher} {World Scientific},\ \bibinfo {address} {Singapore},\ \bibinfo
  {year} {2014})\ Chap.~\bibinfo {chapter} {1}, pp.\ \bibinfo {pages}
  {447--476}\BibitemShut {NoStop}%
\bibitem [{\citenamefont {Chupeau}\ \emph {et~al.}(2020)\citenamefont
  {Chupeau}, \citenamefont {Gladrow}, \citenamefont {Chepelianskii},
  \citenamefont {Keyser},\ and\ \citenamefont
  {Trizac}}]{Chupeau_potential_shape}%
  \BibitemOpen
  \bibfield  {author} {\bibinfo {author} {\bibfnamefont {M.}~\bibnamefont
  {Chupeau}}, \bibinfo {author} {\bibfnamefont {J.}~\bibnamefont {Gladrow}},
  \bibinfo {author} {\bibfnamefont {A.}~\bibnamefont {Chepelianskii}}, \bibinfo
  {author} {\bibfnamefont {U.~F.}\ \bibnamefont {Keyser}},\ and\ \bibinfo
  {author} {\bibfnamefont {E.}~\bibnamefont {Trizac}},\ }\bibfield  {title}
  {\bibinfo {title} {Optimizing brownian escape rates by potential shaping},\
  }\href@noop {} {\bibfield  {journal} {\bibinfo  {journal} {Proceedings of the
  National Academy of Sciences}\ }\textbf {\bibinfo {volume} {117}},\ \bibinfo
  {pages} {1383} (\bibinfo {year} {2020})}\BibitemShut {NoStop}%
\bibitem [{\citenamefont {H\"anggi}\ \emph {et~al.}(1990)\citenamefont
  {H\"anggi}, \citenamefont {Talkner},\ and\ \citenamefont
  {Borkovec}}]{Hanggi_Kramers}%
  \BibitemOpen
  \bibfield  {author} {\bibinfo {author} {\bibfnamefont {P.}~\bibnamefont
  {H\"anggi}}, \bibinfo {author} {\bibfnamefont {P.}~\bibnamefont {Talkner}},\
  and\ \bibinfo {author} {\bibfnamefont {M.}~\bibnamefont {Borkovec}},\
  }\bibfield  {title} {\bibinfo {title} {Reaction-rate theory: fifty years
  after kramers},\ }\href@noop {} {\bibfield  {journal} {\bibinfo  {journal}
  {Rev. Mod. Phys.}\ }\textbf {\bibinfo {volume} {62}},\ \bibinfo {pages} {251}
  (\bibinfo {year} {1990})}\BibitemShut {NoStop}%
\bibitem [{\citenamefont {Grebenkov}(2020)}]{Diffusion_and_surface_phenomena}%
  \BibitemOpen
  \bibfield  {author} {\bibinfo {author} {\bibfnamefont {D.~S.}\ \bibnamefont
  {Grebenkov}},\ }\bibfield  {title} {\bibinfo {title} {Paradigm shift in
  diffusion-mediated surface phenomena},\ }\href@noop {} {\bibfield  {journal}
  {\bibinfo  {journal} {Phys. Rev. Lett.}\ }\textbf {\bibinfo {volume} {125}},\
  \bibinfo {pages} {078102} (\bibinfo {year} {2020})}\BibitemShut {NoStop}%
\bibitem [{\citenamefont {Elber}\ \emph {et~al.}(2020)\citenamefont {Elber},
  \citenamefont {Makarov},\ and\ \citenamefont
  {Orland}}]{Molecular_Kinetics_cond_phase}%
  \BibitemOpen
  \bibfield  {author} {\bibinfo {author} {\bibfnamefont {R.}~\bibnamefont
  {Elber}}, \bibinfo {author} {\bibfnamefont {D.~E.}\ \bibnamefont {Makarov}},\
  and\ \bibinfo {author} {\bibfnamefont {H.}~\bibnamefont {Orland}},\
  }\href@noop {} {\emph {\bibinfo {title} {Molecular Kinetics in Condensed
  Phases}}},\ \bibinfo {edition} {1st}\ ed.\ (\bibinfo  {publisher} {John Wiley
  \& Sons, Incorporated},\ \bibinfo {address} {Newark},\ \bibinfo {year}
  {2020})\BibitemShut {NoStop}%
\bibitem [{\citenamefont {B{\'e}nichou}\ \emph {et~al.}(2010)\citenamefont
  {B{\'e}nichou}, \citenamefont {Chevalier}, \citenamefont {Klafter},
  \citenamefont {Meyer},\ and\ \citenamefont
  {Voituriez}}]{Nature_chem_benichou}%
  \BibitemOpen
  \bibfield  {author} {\bibinfo {author} {\bibfnamefont {O.}~\bibnamefont
  {B{\'e}nichou}}, \bibinfo {author} {\bibfnamefont {C.}~\bibnamefont
  {Chevalier}}, \bibinfo {author} {\bibfnamefont {J.}~\bibnamefont {Klafter}},
  \bibinfo {author} {\bibfnamefont {B.}~\bibnamefont {Meyer}},\ and\ \bibinfo
  {author} {\bibfnamefont {R.}~\bibnamefont {Voituriez}},\ }\bibfield  {title}
  {\bibinfo {title} {Geometry-controlled kinetics},\ }\href@noop {} {\bibfield
  {journal} {\bibinfo  {journal} {Nature Chemistry}\ }\textbf {\bibinfo
  {volume} {2}},\ \bibinfo {pages} {472} (\bibinfo {year} {2010})}\BibitemShut
  {NoStop}%
\bibitem [{\citenamefont {Assaf}\ and\ \citenamefont
  {Meerson}(2010)}]{Asaf2010}%
  \BibitemOpen
  \bibfield  {author} {\bibinfo {author} {\bibfnamefont {M.}~\bibnamefont
  {Assaf}}\ and\ \bibinfo {author} {\bibfnamefont {B.}~\bibnamefont
  {Meerson}},\ }\bibfield  {title} {\bibinfo {title} {Extinction of metastable
  stochastic populations},\ }\href@noop {} {\bibfield  {journal} {\bibinfo
  {journal} {Phys. Rev. E}\ }\textbf {\bibinfo {volume} {81}},\ \bibinfo
  {pages} {021116} (\bibinfo {year} {2010})}\BibitemShut {NoStop}%
\bibitem [{\citenamefont {Ovaskainen}\ and\ \citenamefont
  {Meerson}(2010)}]{Ovaskainen2010}%
  \BibitemOpen
  \bibfield  {author} {\bibinfo {author} {\bibfnamefont {O.}~\bibnamefont
  {Ovaskainen}}\ and\ \bibinfo {author} {\bibfnamefont {B.}~\bibnamefont
  {Meerson}},\ }\bibfield  {title} {\bibinfo {title} {Stochastic models of
  population extinction},\ }\href@noop {} {\bibfield  {journal} {\bibinfo
  {journal} {Trends in Ecology \& Evolution}\ }\textbf {\bibinfo {volume}
  {25}},\ \bibinfo {pages} {643} (\bibinfo {year} {2010})}\BibitemShut
  {NoStop}%
\bibitem [{\citenamefont {Colaiori}(2008)}]{Colaiori2008}%
  \BibitemOpen
  \bibfield  {author} {\bibinfo {author} {\bibfnamefont {F.}~\bibnamefont
  {Colaiori}},\ }\bibfield  {title} {\bibinfo {title} {Exactly solvable model
  of avalanches dynamics for barkhausen crackling noise},\ }\href@noop {}
  {\bibfield  {journal} {\bibinfo  {journal} {Advances in Physics}\ }\textbf
  {\bibinfo {volume} {57}},\ \bibinfo {pages} {287} (\bibinfo {year}
  {2008})}\BibitemShut {NoStop}%
\bibitem [{\citenamefont {Taleb}(2007)}]{taleb2007}%
  \BibitemOpen
  \bibfield  {author} {\bibinfo {author} {\bibfnamefont {N.}~\bibnamefont
  {Taleb}},\ }\href@noop {} {\emph {\bibinfo {title} {The Black Swan: The
  Impact of the Highly Improbable}}},\ Incerto\ (\bibinfo  {publisher} {Random
  House Publishing Group},\ \bibinfo {year} {2007})\BibitemShut {NoStop}%
\bibitem [{\citenamefont {Kundu}\ and\ \citenamefont
  {Reuveni}(2024)}]{Kundu_Preface_2024}%
  \BibitemOpen
  \bibfield  {author} {\bibinfo {author} {\bibfnamefont {A.}~\bibnamefont
  {Kundu}}\ and\ \bibinfo {author} {\bibfnamefont {S.}~\bibnamefont
  {Reuveni}},\ }\bibfield  {title} {\bibinfo {title} {Preface: stochastic
  resetting—theory and applications},\ }\href@noop {} {\bibfield  {journal}
  {\bibinfo  {journal} {Journal of Physics A: Mathematical and Theoretical}\
  }\textbf {\bibinfo {volume} {57}},\ \bibinfo {pages} {060301} (\bibinfo
  {year} {2024})}\BibitemShut {NoStop}%
\bibitem [{\citenamefont {Evans}\ and\ \citenamefont
  {Majumdar}(2011{\natexlab{a}})}]{Satya_and_Evans_PRL}%
  \BibitemOpen
  \bibfield  {author} {\bibinfo {author} {\bibfnamefont {M.~R.}\ \bibnamefont
  {Evans}}\ and\ \bibinfo {author} {\bibfnamefont {S.~N.}\ \bibnamefont
  {Majumdar}},\ }\bibfield  {title} {\bibinfo {title} {Diffusion with
  stochastic resetting},\ }\href@noop {} {\bibfield  {journal} {\bibinfo
  {journal} {Phys. Rev. Lett.}\ }\textbf {\bibinfo {volume} {106}},\ \bibinfo
  {pages} {160601} (\bibinfo {year} {2011}{\natexlab{a}})}\BibitemShut
  {NoStop}%
\bibitem [{\citenamefont {Pal}\ and\ \citenamefont
  {Reuveni}(2017)}]{FPT_reset}%
  \BibitemOpen
  \bibfield  {author} {\bibinfo {author} {\bibfnamefont {A.}~\bibnamefont
  {Pal}}\ and\ \bibinfo {author} {\bibfnamefont {S.}~\bibnamefont {Reuveni}},\
  }\bibfield  {title} {\bibinfo {title} {First passage under restart},\
  }\href@noop {} {\bibfield  {journal} {\bibinfo  {journal} {Phys. Rev. Lett.}\
  }\textbf {\bibinfo {volume} {118}},\ \bibinfo {pages} {030603} (\bibinfo
  {year} {2017})}\BibitemShut {NoStop}%
\bibitem [{\citenamefont {Reuveni}(2016)}]{Optimal_stoc_reset}%
  \BibitemOpen
  \bibfield  {author} {\bibinfo {author} {\bibfnamefont {S.}~\bibnamefont
  {Reuveni}},\ }\bibfield  {title} {\bibinfo {title} {Optimal stochastic
  restart renders fluctuations in first passage times universal},\ }\href@noop
  {} {\bibfield  {journal} {\bibinfo  {journal} {Phys. Rev. Lett.}\ }\textbf
  {\bibinfo {volume} {116}},\ \bibinfo {pages} {170601} (\bibinfo {year}
  {2016})}\BibitemShut {NoStop}%
\bibitem [{\citenamefont {Evans}\ \emph {et~al.}(2020)\citenamefont {Evans},
  \citenamefont {Majumdar},\ and\ \citenamefont {Schehr}}]{stoc_res_and_app}%
  \BibitemOpen
  \bibfield  {author} {\bibinfo {author} {\bibfnamefont {M.~R.}\ \bibnamefont
  {Evans}}, \bibinfo {author} {\bibfnamefont {S.~N.}\ \bibnamefont
  {Majumdar}},\ and\ \bibinfo {author} {\bibfnamefont {G.}~\bibnamefont
  {Schehr}},\ }\bibfield  {title} {\bibinfo {title} {Stochastic resetting and
  applications},\ }\href@noop {} {\bibfield  {journal} {\bibinfo  {journal}
  {Journal of Physics A: Mathematical and Theoretical}\ }\textbf {\bibinfo
  {volume} {53}},\ \bibinfo {pages} {193001} (\bibinfo {year}
  {2020})}\BibitemShut {NoStop}%
\bibitem [{\citenamefont {Evans}\ and\ \citenamefont
  {Majumdar}(2011{\natexlab{b}})}]{diffusion_with_opt_reset}%
  \BibitemOpen
  \bibfield  {author} {\bibinfo {author} {\bibfnamefont {M.~R.}\ \bibnamefont
  {Evans}}\ and\ \bibinfo {author} {\bibfnamefont {S.~N.}\ \bibnamefont
  {Majumdar}},\ }\bibfield  {title} {\bibinfo {title} {Diffusion with optimal
  resetting},\ }\href@noop {} {\bibfield  {journal} {\bibinfo  {journal}
  {Journal of Physics A: Mathematical and Theoretical}\ }\textbf {\bibinfo
  {volume} {44}},\ \bibinfo {pages} {435001} (\bibinfo {year}
  {2011}{\natexlab{b}})}\BibitemShut {NoStop}%
\bibitem [{\citenamefont {Reuveni}\ \emph {et~al.}(2014)\citenamefont
  {Reuveni}, \citenamefont {Urbakh},\ and\ \citenamefont
  {Klafter}}]{sub_unbind}%
  \BibitemOpen
  \bibfield  {author} {\bibinfo {author} {\bibfnamefont {S.}~\bibnamefont
  {Reuveni}}, \bibinfo {author} {\bibfnamefont {M.}~\bibnamefont {Urbakh}},\
  and\ \bibinfo {author} {\bibfnamefont {J.}~\bibnamefont {Klafter}},\
  }\bibfield  {title} {\bibinfo {title} {Role of substrate unbinding in
  michaelis–menten enzymatic reactions},\ }\href@noop {} {\bibfield
  {journal} {\bibinfo  {journal} {Proceedings of the National Academy of
  Sciences}\ }\textbf {\bibinfo {volume} {111}},\ \bibinfo {pages} {4391}
  (\bibinfo {year} {2014})}\BibitemShut {NoStop}%
\bibitem [{\citenamefont {Rotbart}\ \emph {et~al.}(2015)\citenamefont
  {Rotbart}, \citenamefont {Reuveni},\ and\ \citenamefont
  {Urbakh}}]{Michaelis_menten_PRE}%
  \BibitemOpen
  \bibfield  {author} {\bibinfo {author} {\bibfnamefont {T.}~\bibnamefont
  {Rotbart}}, \bibinfo {author} {\bibfnamefont {S.}~\bibnamefont {Reuveni}},\
  and\ \bibinfo {author} {\bibfnamefont {M.}~\bibnamefont {Urbakh}},\
  }\bibfield  {title} {\bibinfo {title} {Michaelis-menten reaction scheme as a
  unified approach towards the optimal restart problem},\ }\href@noop {}
  {\bibfield  {journal} {\bibinfo  {journal} {Phys. Rev. E}\ }\textbf {\bibinfo
  {volume} {92}},\ \bibinfo {pages} {060101} (\bibinfo {year}
  {2015})}\BibitemShut {NoStop}%
\bibitem [{\citenamefont {Chechkin}\ and\ \citenamefont
  {Sokolov}(2018)}]{Sokolov_Renewal}%
  \BibitemOpen
  \bibfield  {author} {\bibinfo {author} {\bibfnamefont {A.}~\bibnamefont
  {Chechkin}}\ and\ \bibinfo {author} {\bibfnamefont {I.~M.}\ \bibnamefont
  {Sokolov}},\ }\bibfield  {title} {\bibinfo {title} {Random search with
  resetting: A unified renewal approach},\ }\href@noop {} {\bibfield  {journal}
  {\bibinfo  {journal} {Phys. Rev. Lett.}\ }\textbf {\bibinfo {volume} {121}},\
  \bibinfo {pages} {050601} (\bibinfo {year} {2018})}\BibitemShut {NoStop}%
\bibitem [{\citenamefont {Blumer}\ \emph {et~al.}(2022)\citenamefont {Blumer},
  \citenamefont {Reuveni},\ and\ \citenamefont
  {Hirshberg}}]{SR_Enhanced_Sampling}%
  \BibitemOpen
  \bibfield  {author} {\bibinfo {author} {\bibfnamefont {O.}~\bibnamefont
  {Blumer}}, \bibinfo {author} {\bibfnamefont {S.}~\bibnamefont {Reuveni}},\
  and\ \bibinfo {author} {\bibfnamefont {B.}~\bibnamefont {Hirshberg}},\
  }\bibfield  {title} {\bibinfo {title} {Stochastic resetting for enhanced
  sampling},\ }\href@noop {} {\bibfield  {journal} {\bibinfo  {journal} {The
  Journal of Physical Chemistry Letters}\ }\textbf {\bibinfo {volume} {13}},\
  \bibinfo {pages} {11230} (\bibinfo {year} {2022})}\BibitemShut {NoStop}%
\bibitem [{\citenamefont {Pal}\ \emph {et~al.}(2020)\citenamefont {Pal},
  \citenamefont {Ku\ifmmode\acute{s}\else\'{s}\fi{}mierz},\ and\ \citenamefont
  {Reuveni}}]{Search_with_home_returns}%
  \BibitemOpen
  \bibfield  {author} {\bibinfo {author} {\bibfnamefont {A.}~\bibnamefont
  {Pal}}, \bibinfo {author} {\bibfnamefont {L.}~\bibnamefont
  {Ku\ifmmode\acute{s}\else\'{s}\fi{}mierz}},\ and\ \bibinfo {author}
  {\bibfnamefont {S.}~\bibnamefont {Reuveni}},\ }\bibfield  {title} {\bibinfo
  {title} {Search with home returns provides advantage under high
  uncertainty},\ }\href@noop {} {\bibfield  {journal} {\bibinfo  {journal}
  {Phys. Rev. Res.}\ }\textbf {\bibinfo {volume} {2}},\ \bibinfo {pages}
  {043174} (\bibinfo {year} {2020})}\BibitemShut {NoStop}%
\bibitem [{\citenamefont {Pal}\ \emph {et~al.}(2019)\citenamefont {Pal},
  \citenamefont {Eliazar},\ and\ \citenamefont
  {Reuveni}}]{Reset_with_branching}%
  \BibitemOpen
  \bibfield  {author} {\bibinfo {author} {\bibfnamefont {A.}~\bibnamefont
  {Pal}}, \bibinfo {author} {\bibfnamefont {I.}~\bibnamefont {Eliazar}},\ and\
  \bibinfo {author} {\bibfnamefont {S.}~\bibnamefont {Reuveni}},\ }\bibfield
  {title} {\bibinfo {title} {First passage under restart with branching},\
  }\href@noop {} {\bibfield  {journal} {\bibinfo  {journal} {Phys. Rev. Lett.}\
  }\textbf {\bibinfo {volume} {122}},\ \bibinfo {pages} {020602} (\bibinfo
  {year} {2019})}\BibitemShut {NoStop}%
\bibitem [{\citenamefont {Bonomo}\ and\ \citenamefont
  {Pal}(2021)}]{StocResetDT}%
  \BibitemOpen
  \bibfield  {author} {\bibinfo {author} {\bibfnamefont {O.~L.}\ \bibnamefont
  {Bonomo}}\ and\ \bibinfo {author} {\bibfnamefont {A.}~\bibnamefont {Pal}},\
  }\bibfield  {title} {\bibinfo {title} {First passage under restart for
  discrete space and time: Application to one-dimensional confined lattice
  random walks},\ }\href@noop {} {\bibfield  {journal} {\bibinfo  {journal}
  {Phys. Rev. E}\ }\textbf {\bibinfo {volume} {103}},\ \bibinfo {pages}
  {052129} (\bibinfo {year} {2021})}\BibitemShut {NoStop}%
\bibitem [{\citenamefont {Feller}(1966)}]{Feller}%
  \BibitemOpen
  \bibfield  {author} {\bibinfo {author} {\bibfnamefont {W.}~\bibnamefont
  {Feller}},\ }\href@noop {} {\emph {\bibinfo {title} {An introduction to
  probability theory and its applications}}},\ \bibinfo {series} {Wiley series
  in probability and mathematical statistics. Probability and mathematical
  statistics}, Vol.~\bibinfo {volume} {2}\ (\bibinfo  {publisher} {J. Wiley},\
  \bibinfo {address} {New York},\ \bibinfo {year} {1966})\BibitemShut {NoStop}%
\bibitem [{\citenamefont {Pal}\ \emph {et~al.}(2022)\citenamefont {Pal},
  \citenamefont {Kostinski},\ and\ \citenamefont
  {Reuveni}}]{inspection_paradox}%
  \BibitemOpen
  \bibfield  {author} {\bibinfo {author} {\bibfnamefont {A.}~\bibnamefont
  {Pal}}, \bibinfo {author} {\bibfnamefont {S.}~\bibnamefont {Kostinski}},\
  and\ \bibinfo {author} {\bibfnamefont {S.}~\bibnamefont {Reuveni}},\
  }\bibfield  {title} {\bibinfo {title} {The inspection paradox in stochastic
  resetting},\ }\href@noop {} {\bibfield  {journal} {\bibinfo  {journal}
  {Journal of Physics A: Mathematical and Theoretical}\ }\textbf {\bibinfo
  {volume} {55}},\ \bibinfo {pages} {021001} (\bibinfo {year}
  {2022})}\BibitemShut {NoStop}%
\bibitem [{\citenamefont {Keidar}\ \emph {et~al.}(2024)\citenamefont {Keidar},
  \citenamefont {Blumer}, \citenamefont {Hirshberg},\ and\ \citenamefont
  {Reuveni}}]{Keidar_Blumer2024}%
  \BibitemOpen
  \bibfield  {author} {\bibinfo {author} {\bibfnamefont {T.~D.}\ \bibnamefont
  {Keidar}}, \bibinfo {author} {\bibfnamefont {O.}~\bibnamefont {Blumer}},
  \bibinfo {author} {\bibfnamefont {B.}~\bibnamefont {Hirshberg}},\ and\
  \bibinfo {author} {\bibfnamefont {S.}~\bibnamefont {Reuveni}},\ }\bibfield
  {title} {\bibinfo {title} {Adaptive resetting for informed search strategies
  and the design of non-equilibrium steady-states},\ }\href@noop {} {\bibfield
  {journal} {\bibinfo  {journal} {arXiv preprint arXiv:2409.14419}\ } (\bibinfo
  {year} {2024})}\BibitemShut {NoStop}%
\bibitem [{\citenamefont {Church}\ \emph {et~al.}(2025)\citenamefont {Church},
  \citenamefont {Blumer}, \citenamefont {Keidar}, \citenamefont {Ploutno},
  \citenamefont {Reuveni},\ and\ \citenamefont
  {Hirshberg}}]{Church_Blumer2025}%
  \BibitemOpen
  \bibfield  {author} {\bibinfo {author} {\bibfnamefont {J.~R.}\ \bibnamefont
  {Church}}, \bibinfo {author} {\bibfnamefont {O.}~\bibnamefont {Blumer}},
  \bibinfo {author} {\bibfnamefont {T.~D.}\ \bibnamefont {Keidar}}, \bibinfo
  {author} {\bibfnamefont {L.}~\bibnamefont {Ploutno}}, \bibinfo {author}
  {\bibfnamefont {S.}~\bibnamefont {Reuveni}},\ and\ \bibinfo {author}
  {\bibfnamefont {B.}~\bibnamefont {Hirshberg}},\ }\bibfield  {title} {\bibinfo
  {title} {Accelerating molecular dynamics through informed resetting},\ }\href
  {https://doi.org/10.1021/acs.jctc.4c01238} {\bibfield  {journal} {\bibinfo
  {journal} {Journal of Chemical Theory and Computation}\ }\textbf {\bibinfo
  {volume} {21}},\ \bibinfo {pages} {605} (\bibinfo {year} {2025})}\BibitemShut
  {NoStop}%
\bibitem [{\citenamefont {Tal-Friedman}\ \emph {et~al.}(2025)\citenamefont
  {Tal-Friedman}, \citenamefont {Keidar}, \citenamefont {Reuveni},\ and\
  \citenamefont {Roichman}}]{Tal_Friedman2025}%
  \BibitemOpen
  \bibfield  {author} {\bibinfo {author} {\bibfnamefont {O.}~\bibnamefont
  {Tal-Friedman}}, \bibinfo {author} {\bibfnamefont {T.~D.}\ \bibnamefont
  {Keidar}}, \bibinfo {author} {\bibfnamefont {S.}~\bibnamefont {Reuveni}},\
  and\ \bibinfo {author} {\bibfnamefont {Y.}~\bibnamefont {Roichman}},\
  }\bibfield  {title} {\bibinfo {title} {Smart resetting: An energy-efficient
  strategy for stochastic search processes},\ }\href@noop {} {\bibfield
  {journal} {\bibinfo  {journal} {Physical Review Research}\ }\textbf {\bibinfo
  {volume} {7}},\ \bibinfo {pages} {013033} (\bibinfo {year}
  {2025})}\BibitemShut {NoStop}%
\bibitem [{\citenamefont {Bonati}\ \emph {et~al.}(2021)\citenamefont {Bonati},
  \citenamefont {Piccini},\ and\ \citenamefont {Parrinello}}]{bonati2021deep}%
  \BibitemOpen
  \bibfield  {author} {\bibinfo {author} {\bibfnamefont {L.}~\bibnamefont
  {Bonati}}, \bibinfo {author} {\bibfnamefont {G.}~\bibnamefont {Piccini}},\
  and\ \bibinfo {author} {\bibfnamefont {M.}~\bibnamefont {Parrinello}},\
  }\bibfield  {title} {\bibinfo {title} {Deep learning the slow modes for rare
  events sampling},\ }\href@noop {} {\bibfield  {journal} {\bibinfo  {journal}
  {Proceedings of the National Academy of Sciences}\ }\textbf {\bibinfo
  {volume} {118}},\ \bibinfo {pages} {e2113533118} (\bibinfo {year}
  {2021})}\BibitemShut {NoStop}%
\bibitem [{\citenamefont {Mendels}\ \emph {et~al.}(2018)\citenamefont
  {Mendels}, \citenamefont {Piccini},\ and\ \citenamefont
  {Parrinello}}]{mendels2018collective}%
  \BibitemOpen
  \bibfield  {author} {\bibinfo {author} {\bibfnamefont {D.}~\bibnamefont
  {Mendels}}, \bibinfo {author} {\bibfnamefont {G.}~\bibnamefont {Piccini}},\
  and\ \bibinfo {author} {\bibfnamefont {M.}~\bibnamefont {Parrinello}},\
  }\bibfield  {title} {\bibinfo {title} {Collective variables from local
  fluctuations},\ }\href@noop {} {\bibfield  {journal} {\bibinfo  {journal}
  {The journal of physical chemistry letters}\ }\textbf {\bibinfo {volume}
  {9}},\ \bibinfo {pages} {2776} (\bibinfo {year} {2018})}\BibitemShut
  {NoStop}%
\bibitem [{\citenamefont {Ray}\ \emph {et~al.}(2019)\citenamefont {Ray},
  \citenamefont {Mondal},\ and\ \citenamefont {Reuveni}}]{RayPeclet}%
  \BibitemOpen
  \bibfield  {author} {\bibinfo {author} {\bibfnamefont {S.}~\bibnamefont
  {Ray}}, \bibinfo {author} {\bibfnamefont {D.}~\bibnamefont {Mondal}},\ and\
  \bibinfo {author} {\bibfnamefont {S.}~\bibnamefont {Reuveni}},\ }\bibfield
  {title} {\bibinfo {title} {Péclet number governs transition to acceleratory
  restart in drift-diffusion},\ }\href@noop {} {\bibfield  {journal} {\bibinfo
  {journal} {Journal of Physics A: Mathematical and Theoretical}\ }\textbf
  {\bibinfo {volume} {52}},\ \bibinfo {pages} {255002} (\bibinfo {year}
  {2019})}\BibitemShut {NoStop}%
\bibitem [{\citenamefont {Plata}\ \emph {et~al.}(2020)\citenamefont {Plata},
  \citenamefont {Gupta},\ and\ \citenamefont {Azaele}}]{Asymmetric_SR}%
  \BibitemOpen
  \bibfield  {author} {\bibinfo {author} {\bibfnamefont {C.~A.}\ \bibnamefont
  {Plata}}, \bibinfo {author} {\bibfnamefont {D.}~\bibnamefont {Gupta}},\ and\
  \bibinfo {author} {\bibfnamefont {S.}~\bibnamefont {Azaele}},\ }\bibfield
  {title} {\bibinfo {title} {Asymmetric stochastic resetting: Modeling
  catastrophic events},\ }\href@noop {} {\bibfield  {journal} {\bibinfo
  {journal} {Phys. Rev. E}\ }\textbf {\bibinfo {volume} {102}},\ \bibinfo
  {pages} {052116} (\bibinfo {year} {2020})}\BibitemShut {NoStop}%
\bibitem [{\citenamefont {Bray}(2000)}]{Bray_XY_Model}%
  \BibitemOpen
  \bibfield  {author} {\bibinfo {author} {\bibfnamefont {A.~J.}\ \bibnamefont
  {Bray}},\ }\bibfield  {title} {\bibinfo {title} {Random walks in logarithmic
  and power-law potentials, nonuniversal persistence, and vortex dynamics in
  the two-dimensional $\mathrm{XY}$ model},\ }\href@noop {} {\bibfield
  {journal} {\bibinfo  {journal} {Phys. Rev. E}\ }\textbf {\bibinfo {volume}
  {62}},\ \bibinfo {pages} {103} (\bibinfo {year} {2000})}\BibitemShut
  {NoStop}%
\bibitem [{\citenamefont {Ray}\ and\ \citenamefont {Reuveni}(2020)}]{RayLog}%
  \BibitemOpen
  \bibfield  {author} {\bibinfo {author} {\bibfnamefont {S.}~\bibnamefont
  {Ray}}\ and\ \bibinfo {author} {\bibfnamefont {S.}~\bibnamefont {Reuveni}},\
  }\bibfield  {title} {\bibinfo {title} {{Diffusion with resetting in a
  logarithmic potential}},\ }\href@noop {} {\bibfield  {journal} {\bibinfo
  {journal} {The Journal of Chemical Physics}\ }\textbf {\bibinfo {volume}
  {152}},\ \bibinfo {pages} {234110} (\bibinfo {year} {2020})}\BibitemShut
  {NoStop}%
\bibitem [{\citenamefont {Elgart}\ \emph {et~al.}(2010)\citenamefont {Elgart},
  \citenamefont {Jia},\ and\ \citenamefont {Kulkarni}}]{Elgart2010}%
  \BibitemOpen
  \bibfield  {author} {\bibinfo {author} {\bibfnamefont {V.}~\bibnamefont
  {Elgart}}, \bibinfo {author} {\bibfnamefont {T.}~\bibnamefont {Jia}},\ and\
  \bibinfo {author} {\bibfnamefont {R.~V.}\ \bibnamefont {Kulkarni}},\
  }\bibfield  {title} {\bibinfo {title} {Applications of little's law to
  stochastic models of gene expression},\ }\href
  {https://doi.org/10.1103/PhysRevE.82.021901} {\bibfield  {journal} {\bibinfo
  {journal} {Phys. Rev. E}\ }\textbf {\bibinfo {volume} {82}},\ \bibinfo
  {pages} {021901} (\bibinfo {year} {2010})}\BibitemShut {NoStop}%
\bibitem [{\citenamefont {Hilfinger}\ \emph {et~al.}(2016)\citenamefont
  {Hilfinger}, \citenamefont {Norman}, \citenamefont {Vinnicombe},\ and\
  \citenamefont {Paulsson}}]{Hilfinger2016}%
  \BibitemOpen
  \bibfield  {author} {\bibinfo {author} {\bibfnamefont {A.}~\bibnamefont
  {Hilfinger}}, \bibinfo {author} {\bibfnamefont {T.~M.}\ \bibnamefont
  {Norman}}, \bibinfo {author} {\bibfnamefont {G.}~\bibnamefont {Vinnicombe}},\
  and\ \bibinfo {author} {\bibfnamefont {J.}~\bibnamefont {Paulsson}},\
  }\bibfield  {title} {\bibinfo {title} {Constraints on fluctuations in
  sparsely characterized biological systems},\ }\href
  {https://doi.org/10.1103/PhysRevLett.116.058101} {\bibfield  {journal}
  {\bibinfo  {journal} {Phys. Rev. Lett.}\ }\textbf {\bibinfo {volume} {116}},\
  \bibinfo {pages} {058101} (\bibinfo {year} {2016})}\BibitemShut {NoStop}%
\bibitem [{\citenamefont {Szavits-Nossan}\ and\ \citenamefont
  {Grima}(2024)}]{Szavits-Nossan2024}%
  \BibitemOpen
  \bibfield  {author} {\bibinfo {author} {\bibfnamefont {J.}~\bibnamefont
  {Szavits-Nossan}}\ and\ \bibinfo {author} {\bibfnamefont {R.}~\bibnamefont
  {Grima}},\ }\bibfield  {title} {\bibinfo {title} {Solving stochastic
  gene-expression models using queueing theory: A tutorial review},\ }\href
  {https://doi.org/10.1016/j.bpj.2024.04.004} {\bibfield  {journal} {\bibinfo
  {journal} {Biophysical Journal}\ }\textbf {\bibinfo {volume} {123}},\
  \bibinfo {pages} {1034} (\bibinfo {year} {2024})}\BibitemShut {NoStop}%
\bibitem [{\citenamefont {Little}(1961)}]{Little_1961}%
  \BibitemOpen
  \bibfield  {author} {\bibinfo {author} {\bibfnamefont {J.~D.~C.}\
  \bibnamefont {Little}},\ }\bibfield  {title} {\bibinfo {title} {A proof for
  the queuing formula: L = $\lambda$ w},\ }\href
  {https://doi.org/10.1287/opre.9.3.383} {\bibfield  {journal} {\bibinfo
  {journal} {Operations Research}\ }\textbf {\bibinfo {volume} {9}},\ \bibinfo
  {pages} {383} (\bibinfo {year} {1961})},\ \Eprint
  {https://arxiv.org/abs/https://doi.org/10.1287/opre.9.3.383}
  {https://doi.org/10.1287/opre.9.3.383} \BibitemShut {NoStop}%
\bibitem [{\citenamefont {Moffitt}\ and\ \citenamefont
  {Bustamante}(2014)}]{Moffitt2014}%
  \BibitemOpen
  \bibfield  {author} {\bibinfo {author} {\bibfnamefont {J.~R.}\ \bibnamefont
  {Moffitt}}\ and\ \bibinfo {author} {\bibfnamefont {C.}~\bibnamefont
  {Bustamante}},\ }\bibfield  {title} {\bibinfo {title} {Extracting signal from
  noise: kinetic mechanisms from a michaelis–menten-like expression for
  enzymatic fluctuations},\ }\href
  {https://doi.org/https://doi.org/10.1111/febs.12545} {\bibfield  {journal}
  {\bibinfo  {journal} {The FEBS Journal}\ }\textbf {\bibinfo {volume} {281}},\
  \bibinfo {pages} {498} (\bibinfo {year} {2014})},\ \Eprint
  {https://arxiv.org/abs/https://febs.onlinelibrary.wiley.com/doi/pdf/10.1111/febs.12545}
  {https://febs.onlinelibrary.wiley.com/doi/pdf/10.1111/febs.12545}
  \BibitemShut {NoStop}%
\bibitem [{\citenamefont {David}\ and\ \citenamefont
  {Larry}(1987)}]{Aldous_Shepp1987}%
  \BibitemOpen
  \bibfield  {author} {\bibinfo {author} {\bibfnamefont {A.}~\bibnamefont
  {David}}\ and\ \bibinfo {author} {\bibfnamefont {S.}~\bibnamefont {Larry}},\
  }\bibfield  {title} {\bibinfo {title} {The least variable phase type
  distribution is erlang},\ }\href@noop {} {\bibfield  {journal} {\bibinfo
  {journal} {Stochastic Models}\ }\textbf {\bibinfo {volume} {3}},\ \bibinfo
  {pages} {467} (\bibinfo {year} {1987})}\BibitemShut {NoStop}%
\bibitem [{\citenamefont {Moffitt}\ \emph {et~al.}(2010)\citenamefont
  {Moffitt}, \citenamefont {Chemla},\ and\ \citenamefont
  {Bustamante}}]{Moffitt2010PNAS}%
  \BibitemOpen
  \bibfield  {author} {\bibinfo {author} {\bibfnamefont {J.~R.}\ \bibnamefont
  {Moffitt}}, \bibinfo {author} {\bibfnamefont {Y.~R.}\ \bibnamefont
  {Chemla}},\ and\ \bibinfo {author} {\bibfnamefont {C.}~\bibnamefont
  {Bustamante}},\ }\bibfield  {title} {\bibinfo {title} {Mechanistic
  constraints from the substrate concentration dependence of enzymatic
  fluctuations},\ }\href {https://doi.org/10.1073/pnas.1006997107} {\bibfield
  {journal} {\bibinfo  {journal} {Proceedings of the National Academy of
  Sciences}\ }\textbf {\bibinfo {volume} {107}},\ \bibinfo {pages} {15739}
  (\bibinfo {year} {2010})},\ \Eprint
  {https://arxiv.org/abs/https://www.pnas.org/doi/pdf/10.1073/pnas.1006997107}
  {https://www.pnas.org/doi/pdf/10.1073/pnas.1006997107} \BibitemShut {NoStop}%
\bibitem [{\citenamefont {Schnitzer}\ and\ \citenamefont
  {Block}(1997)}]{Schnitzer1997}%
  \BibitemOpen
  \bibfield  {author} {\bibinfo {author} {\bibfnamefont {M.~J.}\ \bibnamefont
  {Schnitzer}}\ and\ \bibinfo {author} {\bibfnamefont {S.~M.}\ \bibnamefont
  {Block}},\ }\bibfield  {title} {\bibinfo {title} {Kinesin hydrolyses one atp
  per 8-nm step},\ }\href {https://doi.org/10.1038/41111} {\bibfield  {journal}
  {\bibinfo  {journal} {Nature}\ }\textbf {\bibinfo {volume} {388}},\ \bibinfo
  {pages} {386} (\bibinfo {year} {1997})}\BibitemShut {NoStop}%
\bibitem [{\citenamefont {Keilson}(1979)}]{Keilson1979}%
  \BibitemOpen
  \bibfield  {author} {\bibinfo {author} {\bibfnamefont {J.}~\bibnamefont
  {Keilson}},\ }\href@noop {} {\emph {\bibinfo {title} {Markov chain
  models--rarity and exponentiality}}},\ Applied mathematical sciences ; v.
  028\ (\bibinfo  {publisher} {Springer-Verlag},\ \bibinfo {address} {New
  York},\ \bibinfo {year} {1979})\BibitemShut {NoStop}%
\bibitem [{\citenamefont {Satija}\ \emph {et~al.}(2020)\citenamefont {Satija},
  \citenamefont {Berezhkovskii},\ and\ \citenamefont {Makarov}}]{Satija2020}%
  \BibitemOpen
  \bibfield  {author} {\bibinfo {author} {\bibfnamefont {R.}~\bibnamefont
  {Satija}}, \bibinfo {author} {\bibfnamefont {A.~M.}\ \bibnamefont
  {Berezhkovskii}},\ and\ \bibinfo {author} {\bibfnamefont {D.~E.}\
  \bibnamefont {Makarov}},\ }\bibfield  {title} {\bibinfo {title} {Broad
  distributions of transition-path times are fingerprints of
  multidimensionality of the underlying free energy landscapes},\ }\href
  {https://doi.org/10.1073/pnas.2008307117} {\bibfield  {journal} {\bibinfo
  {journal} {Proceedings of the National Academy of Sciences}\ }\textbf
  {\bibinfo {volume} {117}},\ \bibinfo {pages} {27116} (\bibinfo {year}
  {2020})},\ \Eprint
  {https://arxiv.org/abs/https://www.pnas.org/doi/pdf/10.1073/pnas.2008307117}
  {https://www.pnas.org/doi/pdf/10.1073/pnas.2008307117} \BibitemShut {NoStop}%
\bibitem [{\citenamefont {Meir}\ \emph {et~al.}(2025)\citenamefont {Meir},
  \citenamefont {Keidar}, \citenamefont {Reuveni},\ and\ \citenamefont
  {Hirshberg}}]{Meir_2025}%
  \BibitemOpen
  \bibfield  {author} {\bibinfo {author} {\bibfnamefont {S.}~\bibnamefont
  {Meir}}, \bibinfo {author} {\bibfnamefont {T.~D.}\ \bibnamefont {Keidar}},
  \bibinfo {author} {\bibfnamefont {S.}~\bibnamefont {Reuveni}},\ and\ \bibinfo
  {author} {\bibfnamefont {B.}~\bibnamefont {Hirshberg}},\ }\bibfield  {title}
  {\bibinfo {title} {First-passage approach to optimizing perturbations for
  improved training of machine learning models},\ }\href
  {https://doi.org/10.1088/2632-2153/add8df} {\bibfield  {journal} {\bibinfo
  {journal} {Machine Learning: Science and Technology}\ }\textbf {\bibinfo
  {volume} {6}},\ \bibinfo {pages} {025053} (\bibinfo {year}
  {2025})}\BibitemShut {NoStop}%
\bibitem [{\citenamefont {Istratov}\ and\ \citenamefont
  {Vyvenko}(1999)}]{Istratov1999}%
  \BibitemOpen
  \bibfield  {author} {\bibinfo {author} {\bibfnamefont {A.~A.}\ \bibnamefont
  {Istratov}}\ and\ \bibinfo {author} {\bibfnamefont {O.~F.}\ \bibnamefont
  {Vyvenko}},\ }\bibfield  {title} {\bibinfo {title} {Exponential analysis in
  physical phenomena},\ }\href {https://doi.org/10.1063/1.1149581} {\bibfield
  {journal} {\bibinfo  {journal} {Review of Scientific Instruments}\ }\textbf
  {\bibinfo {volume} {70}},\ \bibinfo {pages} {1233} (\bibinfo {year}
  {1999})},\ \Eprint
  {https://arxiv.org/abs/https://pubs.aip.org/aip/rsi/article-pdf/70/2/1233/19237413/1233\_1\_online.pdf}
  {https://pubs.aip.org/aip/rsi/article-pdf/70/2/1233/19237413/1233\_1\_online.pdf}
  \BibitemShut {NoStop}%
\bibitem [{\citenamefont {Golub}\ \emph {et~al.}(1979)\citenamefont {Golub},
  \citenamefont {Heath},\ and\ \citenamefont {Wahba}}]{Golub1979}%
  \BibitemOpen
  \bibfield  {author} {\bibinfo {author} {\bibfnamefont {G.~H.}\ \bibnamefont
  {Golub}}, \bibinfo {author} {\bibfnamefont {M.}~\bibnamefont {Heath}},\ and\
  \bibinfo {author} {\bibfnamefont {G.}~\bibnamefont {Wahba}},\ }\bibfield
  {title} {\bibinfo {title} {Generalized cross-validation as a method for
  choosing a good ridge parameter},\ }\href@noop {} {\bibfield  {journal}
  {\bibinfo  {journal} {Technometrics}\ }\textbf {\bibinfo {volume} {21}},\
  \bibinfo {pages} {215} (\bibinfo {year} {1979})}\BibitemShut {NoStop}%
\end{thebibliography}%

\end{document}